\documentclass[pdflatex,sn-mathphys-num]{sn-jnl}

\usepackage{graphicx}%
\usepackage{multirow}%
\usepackage{amsmath,amssymb,amsfonts}%
\usepackage{amsthm}%
\usepackage{mathrsfs}%
\usepackage[title]{appendix}%
\usepackage{xcolor}%
\usepackage{textcomp}%
\usepackage{manyfoot}%
\usepackage{booktabs}%
\usepackage{algorithm}%
\usepackage{algorithmicx}%
\usepackage{algpseudocode}%
\usepackage{listings}%

\usepackage{tabularx}%
\usepackage{amsmath}%
\usepackage{ragged2e}%
\usepackage{array}%
\usepackage{makecell}%
\usepackage{subcaption}%
\usepackage{adjustbox}%

\theoremstyle{thmstyleone}%
\theoremstyle{thmstyletwo}%

\theoremstyle{thmstylethree}%

\begin{document}

\title[Article Title]{Towards Secure Retrieval-Augmented Generation: A Comprehensive Review of Threats, Defenses and Benchmarks}


\author[1,2]{\fnm{Yanming} \sur{Mu}}\email{mu242211@163.com}

\author*[1,2]{\fnm{Hao} \sur{Hu}}\email{wjjhh\_908@163.com}

\author[1,2]{\fnm{Feiyang} \sur{Li}}\email{lfyxxgcdx@163.com}

\author[3]{\fnm{Qiao} \sur{Yuan}}\email{13883065823@163.com}

\author[3]{\fnm{Jiang} \sur{Wu}}\email{liam181113@163.com}

\author[3]{\fnm{Zichun} \sur{Liu}}\email{ieuliuzc@163.com}

\author[3]{\fnm{Pengcheng} \sur{Liu}}\email{liupengcheng2016@163.com}

\author[3]{\fnm{Mei} \sur{Wang}}\email{cybersecuritys@126.com}

\author[3]{\fnm{Hongwei} \sur{Zhou}}\email{hong\_wei\_zhou@126.com}

\author[4]{\fnm{Yuling} \sur{Liu}}\email{liuyuling@iie.ac.cn}

\affil[1]{\orgname{State Key Laboratory of Mathematical Engineering and Advanced Computing},  \orgaddress{\city{Zhengzhou}, \postcode{450001},  \country{China}}}

\affil[2]{\orgname{Information Engineering University}, \orgaddress{\city{Zhengzhou}, \postcode{450001}, \country{China}}}

\affil[3]{\orgname{Henan Key Laboratory of Information Security}, \orgaddress{\city{Zhengzhou}, \postcode{450001},  \country{China}}}

\affil[4]{\orgdiv{School of Cyber Security}, \orgname{University of Chinese Academy of Sciences}, \orgaddress{\city{Beijing}, \postcode{100000}, \country{China}}}

\abstract{Retrieval-Augmented Generation (RAG) significantly mitigates the hallucinations and domain knowledge deficiency in large language models by incorporating external knowledge bases. However, the multi-module architecture of RAG introduces complex system-level security vulnerabilities. Guided by the RAG workflow, this paper analyzes the underlying vulnerability mechanisms and systematically categorizes core threat vectors such as data poisoning, adversarial attacks, and membership inference attacks. Based on this threat assessment, we construct a taxonomy of RAG defense technologies from a dual perspective encompassing both input and output stages. The input-side analysis reviews data protection mechanisms including dynamic access control, homomorphic encryption retrieval, and adversarial pre-filtering. The output-side examination summarizes advanced leakage prevention techniques such as federated learning isolation, differential privacy perturbation, and lightweight data sanitization. To establish a unified benchmark for future experimental design, we consolidate authoritative test datasets, security standards, and evaluation frameworks. To the best of our knowledge, this paper presents the first end-to-end survey dedicated to the security of RAG systems. Distinct from existing literature that isolates specific vulnerabilities, we systematically map the entire pipeline—providing a unified analysis of threat models, defense mechanisms, and evaluation benchmarks. By enabling deep insights into potential risks, this work seeks to foster the development of highly robust and trustworthy next-generation RAG systems.}




\maketitle

\section{Introduction}\label{sec1}

 The proposal of the Transformer architecture in 2017 \citep{1} revolutionized the field of Natural Language Processing (NLP) and laid the core technical foundation for the rapid development of Large Language Models (LLMs). From early milestones such as BERT \citep{2} and GPT \citep{3} to subsequent iterations like GPT-4 \citep{4} and Deepseek-R1 \citep{5}, the technical evolution of LLMs continues to drive breakthroughs in related research and applications. Currently, these models demonstrate strong application potential across various interdisciplinary scenarios, including medicine \citep{6}, chemical research \citep{7}, energy system optimization \citep{8}, and robotic entity control \citep{9}. This progress has not only expanded the technical boundaries of NLP but also provided new technical pathways for solving complex problems.

Compared to smaller language models, LLMs possess unique emergent abilities \citep{10}, such as in-context learning \citep{11} and logical reasoning \citep{12}. Furthermore, supported by massive training data, pre-trained LLMs accumulate vast amounts of world knowledge. These advantages have enabled the application of language models to expand from traditional language modeling to practical task solving, covering fundamental tasks like text classification and sentiment analysis, as well as more challenging areas such as high-level task planning and complex decision-making.

However, the performance of LLMs is constrained by the quality and scope of their training data. For instance, when presented with real-time or domain-specific queries, the underlying LLM may generate answers that appear plausible but are factually incorrect. This phenomenon is known as hallucination \citep{13}.

To address the hallucination issue in LLMs, the Facebook (now Meta) AI Research team \citep{14} introduced Retrieval-Augmented Generation (RAG) in 2020. By integrating retrieval and generation modules, RAG aims to improve the timeliness and accuracy of generated content, effectively mitigating common challenges such as hallucinations \citep{13} and knowledge obsolescence \citep{15,16,19}. Following the validation of its efficacy, RAG entered a period of rapid development. The proposal of technical routes such as GraphRAG \citep{20} and AgenticRAG \citep{21} has significantly broadened the capabilities of RAG. This ability to incorporate external knowledge sources has demonstrated immense potential in fields such as Question Answering (QA) systems \citep{22}, medical applications \citep{23}, and academic education services \citep{24}. For example, the introduction of the RAGCare-QA dataset aims to evaluate the performance of RAG pipelines in theoretical medical QA within medical education \citep{25}. Despite the broad prospects of RAG technology, its multi-modular architecture introduces complex security and privacy risks.

The widespread application of RAG frameworks has prompted in-depth academic investigation into their security characteristics. Studies indicate that RAG may, in certain cases, compromise model security and alter its security profile \citep{35}. The potential attack surface of RAG includes data poisoning attacks \citep{36}, membership inference attacks \citep{37}, and adversarial attacks \citep{38}. For instance, data poisoning attacks can manipulate system outputs by injecting a small amount of malicious text into the knowledge base \citep{16}. Membership inference attacks allow attackers to infer data entries within the system's database based on system outputs \citep{37}. Additionally, adversarial attacks can manipulate system outputs by adding minor perturbations to inputs to bypass detection mechanisms \citep{38}.

To address these emerging security challenges, researchers from top universities and companies worldwide—including the University of Oxford, Amazon AWS AI, the National University of Singapore, the University of Cambridge, the Institute of Information Engineering (CAS), and Tsinghua University—are actively exploring various defense strategies and frameworks. Their findings have been published in top-tier venues such as SIGIR, USENIX Security Symposium, CCS, NDSS, EMNLP, ACL, and TPAMI. For example, \cite{41} aim to enhance the privacy and security of RAG applications through privacy-aware retrieval mechanisms, decentralized access control, and real-time model auditing.\cite{27} proposed a risk assessment and mitigation framework to ensure security when integrating sensitive data in RAG. Furthermore, as RAG becomes increasingly prevalent in industry, data and service security have become critical priorities \citep{27}. These efforts collectively advance the field of RAG security and highlight the urgency of comprehensive security analysis and defense research.

\begin{table*}[t]
  \centering
  \small 
  \caption{Comparison of Research Scopes in Existing Survey Literature on RAG Security 
           (\checkmark{} indicates the existence of this dimension; $\times$ indicates the absence of this dimension)}
  \label{tab:rag_security_survey}
  
  \begin{tabularx}{\textwidth}{@{}l*{6}{>{\centering\arraybackslash}X}@{}}
    \toprule
    \textbf{Aspect} & \textbf{This survey} & \textbf{Wu (2025)} & \textbf{Gu (2025)} & \textbf{He (2025)} & \textbf{Arz (2025)} & \textbf{Wang (2025)} \\
    \midrule
    RAG Architecture Anatomy & \checkmark & \checkmark & \checkmark & $\times$ & \checkmark & \checkmark \\
    Attack Vectors \& Threat Surfaces & \checkmark & $\times$ & \checkmark & \checkmark & \checkmark & \checkmark \\
    Security Defense \& Mitigation Strategies & \checkmark & $\times$ & \checkmark & \checkmark & $\times$ & \checkmark \\
    Evaluation Benchmarks \& Metrics System & \checkmark & \checkmark & $\times$ & $\times$ & $\times$ & $\times$ \\
    \bottomrule
  \end{tabularx}
\end{table*}

As illustrated in Table \ref{tab:rag_security_survey}, existing survey literature on RAG security presents an incomplete research scope. This survey achieves comprehensive coverage across four core dimensions, specifically encompassing architectural analysis, threat analysis, defense strategies, and evaluation metrics.

This survey aims to provide a systematic review and in-depth analysis of existing research in the field of RAG security, covering threat models, attack types, potential vulnerabilities, and proposed defense strategies. To gain a deep understanding of the latest developments in RAG security research, we conducted a systematic and extensive survey of the relevant literature. Based on a comprehensive analysis of 152 relevant papers, we examine the security risks introduced by each module within the RAG multi-modular architecture and compare the characteristics and impacts of different attack vectors. Specifically, this survey will:
\begin{enumerate}
    \item Outline the architecture and core components of RAG: Clarify how the retrieval and generation modules collaborate and their respective significance in security \citep{17,19}.
    \item Identify and categorize major security threats: Analyze attack types such as knowledge poisoning, adversarial attacks, and membership inference attacks, and explore how these attacks exploit specific mechanisms of RAG to achieve malicious purposes \citep{1,16}.
    \item Review existing security assessment methods and tools: Examine current methodologies for evaluating RAG security, such as comparing security performance between RAG and non-RAG frameworks \citep{35}.
    \item Discuss mitigation strategies and defense mechanisms: Summarize proposed techniques, including privacy-enhancing technologies, robustness defenses, and countermeasures against specific attacks \citep{27,41}.
    \item Propose future research directions: Identify current limitations and suggest future focus areas, exploring the integration of searchable encryption with RAG security protection to build more secure and reliable RAG technologies.
\end{enumerate}
By providing a comprehensive overview of the RAG security landscape, this survey aims to serve as a valuable reference for researchers, developers, and policymakers. It seeks to facilitate a better understanding of and response to the security and privacy challenges posed by RAG in practical applications, thereby promoting the healthy development and widespread adoption of the technology.
\begin{figure}[!htbp]
    \centering
    \begin{subfigure}{0.48\textwidth}
        \centering
        \includegraphics[width=1\textwidth]{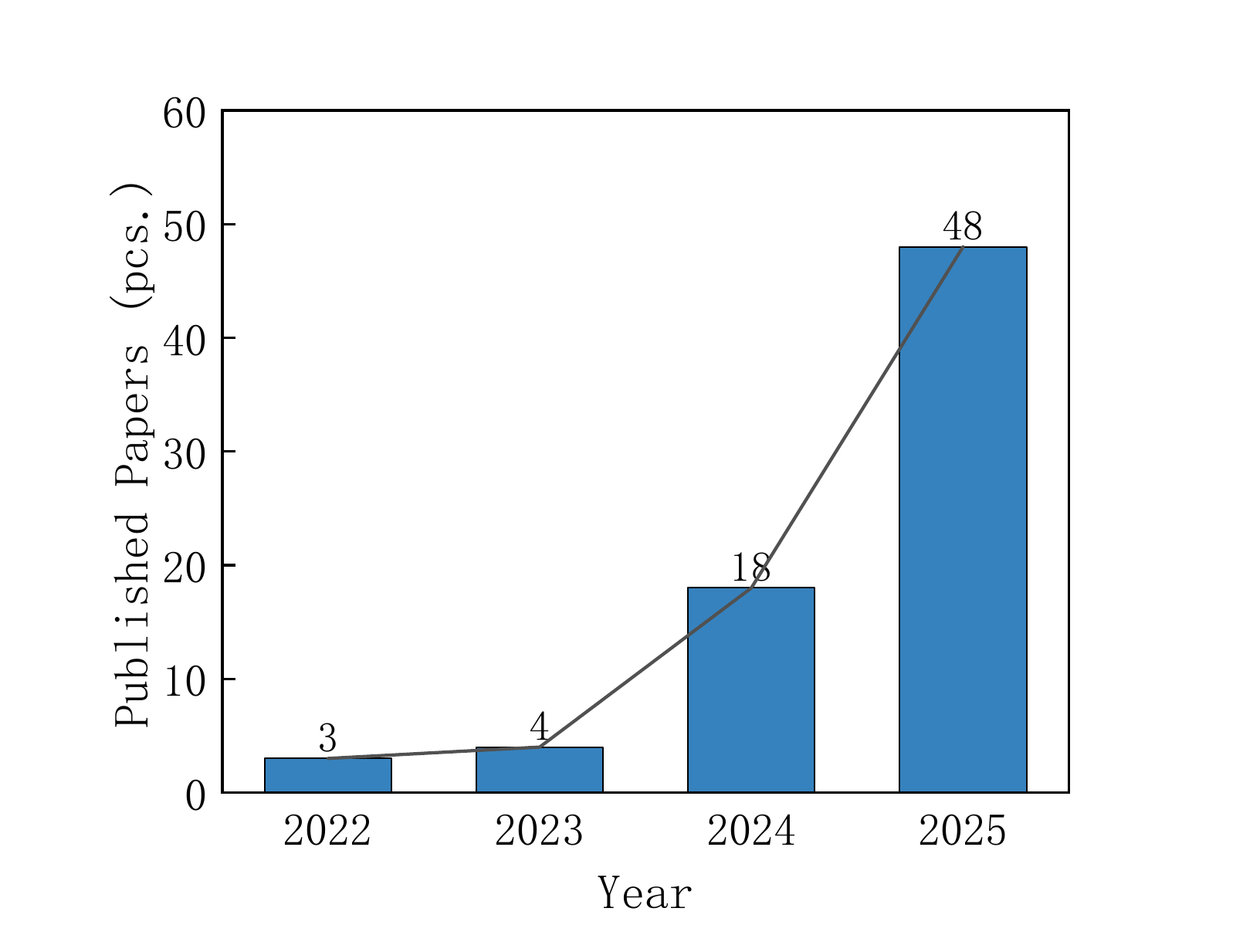}
        \caption{The exponential increase in RAG security research from 2022 to 2025. This figure highlights the dramatic shift in academic attention toward the security aspects of Retrieval-Augmented Generation. While the topic received minimal attention prior to 2024 (averaging fewer than 5 papers annually), the publication count surged significantly to 18 in 2024 and further skyrocketed to 48 in 2025. The steep trend line clearly indicates that RAG security has rapidly transitioned from a niche topic into a mainstream, critical research frontier within the broader large language model (LLM) community.}
        \label{fig:1}
    \end{subfigure}
    \hfill
    \begin{subfigure}{0.48\textwidth}
        \centering
        \includegraphics[width=1\textwidth]{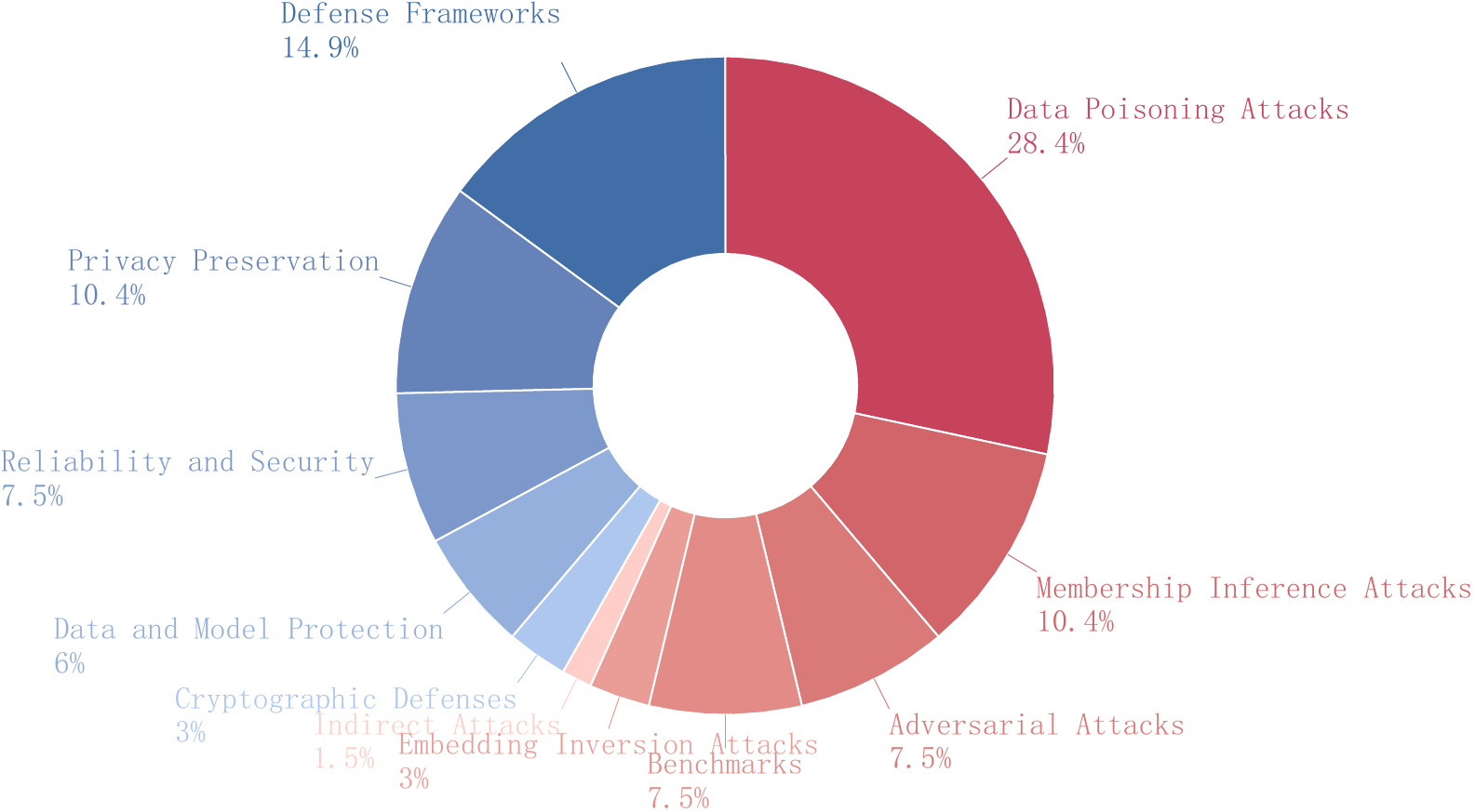}
        \caption{Composition of RAG security literature by specific threat and protection categories. The chart demonstrates a clear focus on offensive research within the current academic landscape. ``Data Poisoning Attacks" emerges as the most prominent research area (28.4\%), significantly overshadowing other specific attack and defense vectors. Defensive studies, indicated by the blue slices, are led by generalized ``Defense Frameworks" (14.9\%) and ``Privacy Preservation" (10.4\%). Overall, the chart visualizes the current phase of RAG security research, which is still primarily heavily concentrated on identifying diverse attack surfaces rather than consolidating unified defenses.}
        \label{fig:2}
    \end{subfigure}
    \caption{Overview of RAG security research trends and literature composition}
    \label{fig:1_2}
\end{figure}
\begin{figure}[!htbp]
    \centering 
    \includegraphics[width=1\textwidth]{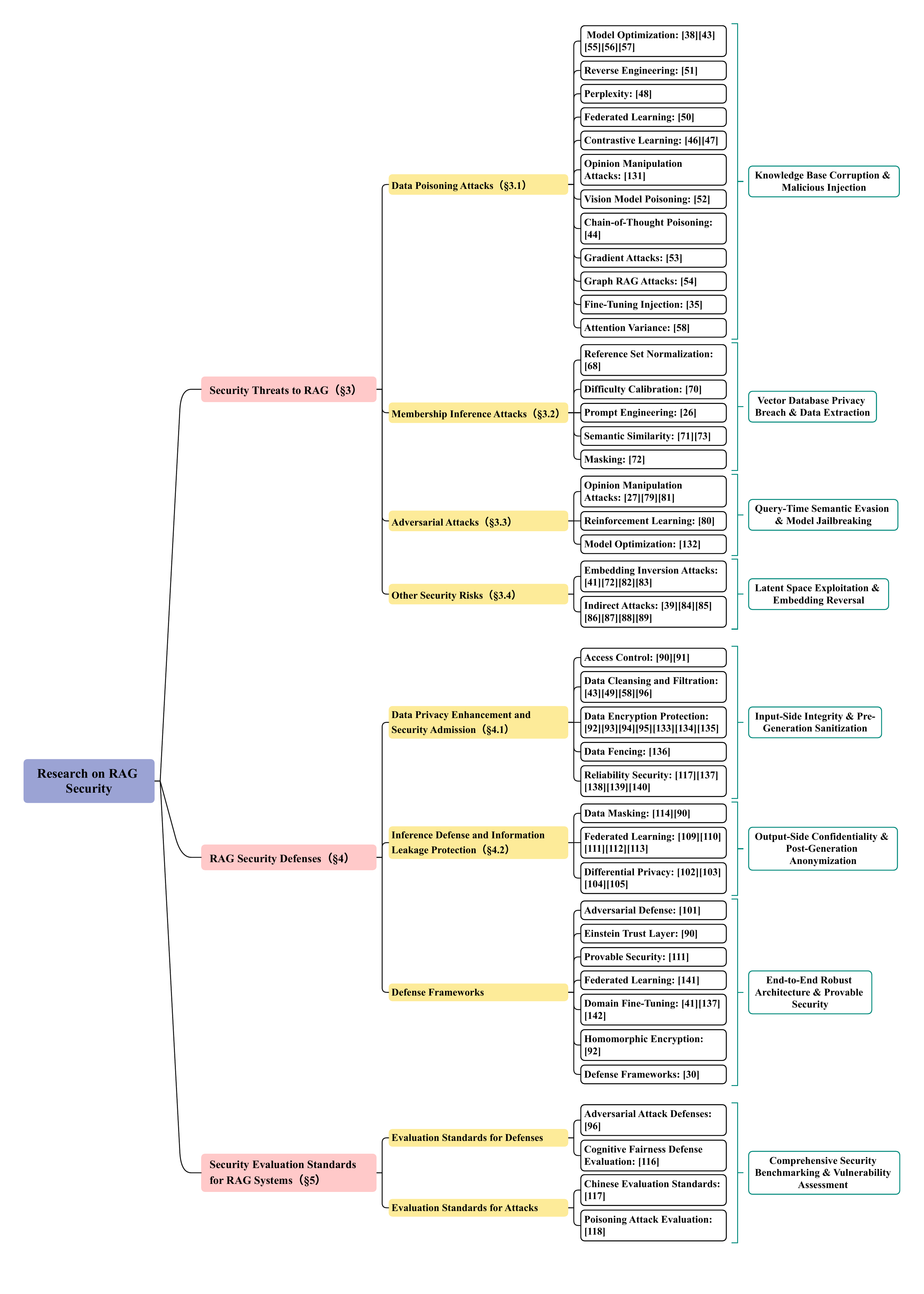}
    \caption{Structured taxonomy mapping the landscape of RAG security research. The diagram organizes the field into Threats, Defenses, and Evaluation Standards, cascading down to specific methodologies and key literature citations. To provide architectural context, the right-hand column groups these methodologies based on where they operate within the RAG pipeline, differentiating between attacks targeting specific modules (e.g., Retrievers) and defenses applied at specific stages (e.g., Full-Pipeline Defenses).}
    \label{fig:3}
\end{figure}
\section{Overview of RAG Technology}\label{sec2}

In 2020, the Facebook (now Meta) AI Research team \citep{14} introduced Retrieval-Augmented Generation (RAG). Early research often defined it as ``external memory" or ``external knowledge bases." Although there was initially debate within the industry regarding the choice between RAG and Fine-tuning, RAG ultimately established its indispensable status within the AI ecosystem due to its significant cost-effectiveness and real-time advantages \citep{28}. By early 2024, the maturation of LLMOps architectures significantly lowered the barriers to system construction, marking the entry of RAG into a stage of rapid development. With the performance of open-source Large Language Models (LLMs) approaching that of commercial closed-source models and breakthroughs in Long Context technology, the foundational applications of RAG have become popularized. Furthermore, the dialectical relationship between Long Context technology and RAG—characterized by synergy and complementarity—has been clarified in both theory and practice \citep{29}.

To address three core challenges—difficulty in processing unstructured data, low recall rates, and the semantic gap—a number of breakthrough technologies have emerged in the field. In terms of document parsing, multi-modal parsing tools such as DeepDoc, MinerU, and Docling have risen to prominence, driving document intelligence technologies based on generative architectures to gradually replace traditional computer vision models. Regarding retrieval strategies, hybrid search modes fusing dense vectors, sparse vectors, and full-text retrieval have become the mainstream paradigm, causing the independent value of pure vector databases to gradually decline \citep{29}. To effectively bridge the semantic gap, GraphRAG \citep{20} and its derivative architectures (including FastGraphRAG, LightRAG, and LazyGraphRAG) have improved system understanding by constructing knowledge graphs and reinforcing entity associations. Meanwhile, research such as RAPTOR and SiReRAG \citep{31} has further optimized recall performance in multi-hop QA and fuzzy query scenarios \citep{30}.

In terms of ranking mechanisms, model architectures are evolving from traditional Cross-Encoders to Late Interaction models based on tensors, such as ColBERT and ColPali. Databases like Infinity and Vespa provide native tensor support, achieving an effective balance between ranking precision and computational cost \citep{32}. Additionally, Agentic RAG has become a focal point in the industry. Frameworks like LangGraph endow systems with closed-loop reflection capabilities, which, combined with reasoning frameworks such as Multi-Agent collaboration and Monte Carlo Tree Search (MCTS), effectively expand the boundaries of processing in complex scenarios. The deep integration of Memory management with RAG has also become a key evolutionary direction \citep{33,34}. With the deep iteration of Vision-Language Models (VLMs) like GPT-4o and PaliGemma, Multi-modal RAG is rising rapidly, forming parallel technical pathways of direct vector generation and generalized OCR-to-text conversion \citep{32}. Simultaneously, the application of data cleaning technologies such as Late Chunking and Contextual Chunking continues to improve the quality of data ingestion. Collectively, these technical advancements are driving the deep implementation of RAG technology into complex enterprise-level application scenarios.

This section provides an overview of Retrieval-Augmented Generation technology. The core objective is to deconstruct its complete technical workflow and deeply analyze the inherent security mechanism defects within each stage, laying a solid theoretical and technical foundation for the subsequent classification of security threats, discussion of defense technologies, and proposal of relevant solutions. The general technical workflow of RAG is outlined below.

\begin{figure}[!htbp]
    \centering
    \includegraphics[width=1\textwidth]{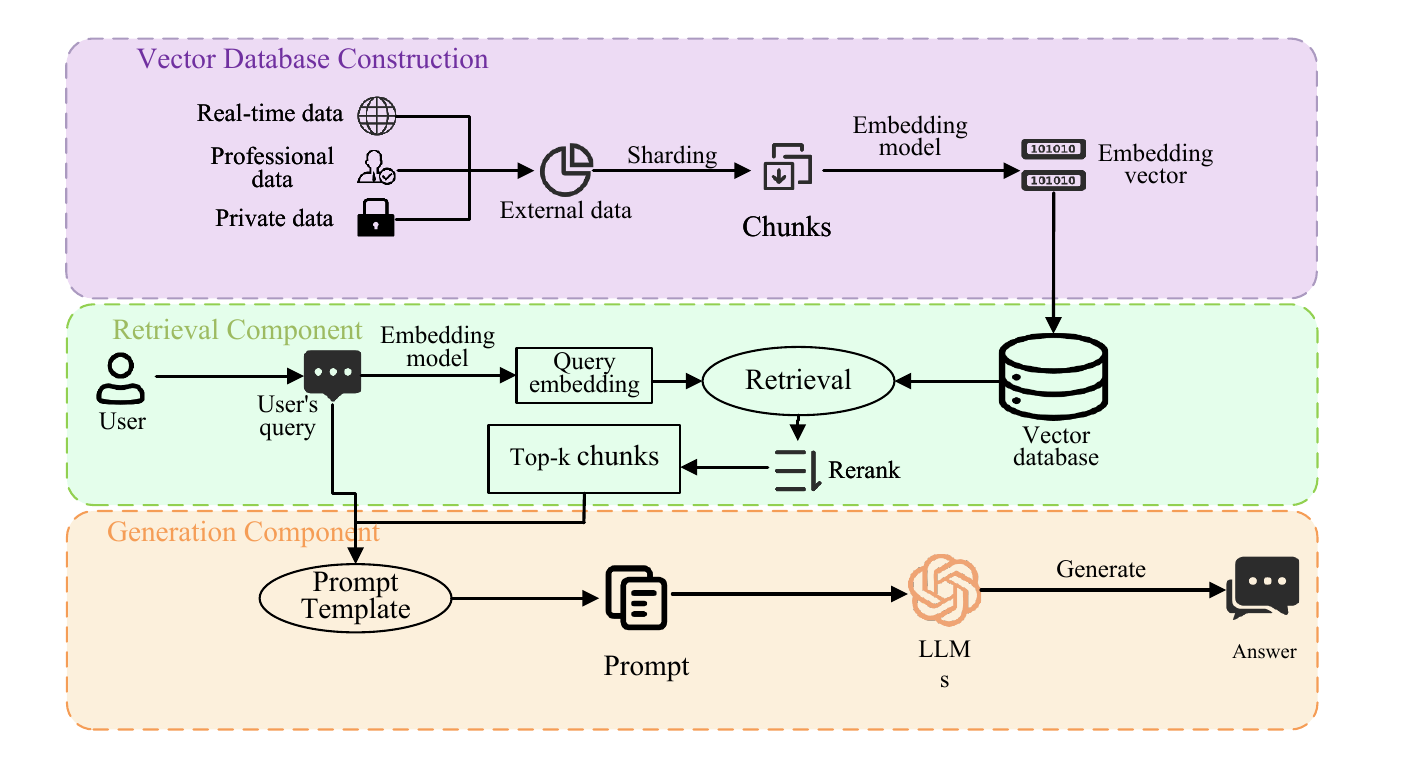}
    \caption{RAG technical workflow: i) Vector database construction, which involves calculating semantic vectors for data chunks via data chunking and embedding models to establish the vector database ; ii) Retriever, responsible for retrieving the top-k data chunks most relevant to the user query from the database ; iii) Generator, responsible for integrating the top-k data chunks with the user query and submitting them to the large language model for response generation .
}
    \label{fig:4}
\end{figure}

As shown in Figure \ref{fig:4}, the technical workflow of RAG primarily consists of three core modules: Vector Database Construction, the Retriever, and the Generator.

Vector Database Construction Phase: This phase aims to transform external knowledge into a retrievable vector index. Unlike the pre-training data of general LLMs, the external knowledge base of RAG emphasizes specificity, real-time availability, and privacy. The database construction process mainly involves two key steps: Chunking and Embedding.

Chunking: This step is responsible for slicing multi-dimensional and multi-modal data into semantic units suitable for model processing. The chunking strategy requires a balance in granularity: chunks that are too large may cause the model's attention to disperse, affecting the extraction of central concepts within the text block; conversely, chunks that are too small may disrupt the semantic integrity of the text.
Embedding: Building on chunking, an embedding model is used to encode data segments, mapping them into high-dimensional vectors. Data segments processed in this manner possess advanced semantic representation capabilities, establishing a foundation for improving the accuracy and relevance of subsequent retrieval tasks.

The Retriever: The Retriever is a critical component of RAG, with the goal of identifying and recalling the Top-$k$ data chunks most semantically relevant to the user's query. To balance efficiency and accuracy, RAG employs a dual mechanism of retrieval followed by re-ranking to recall data segments. First, the Retriever embeds the user's query to obtain its vector representation. Subsequently, utilizing similarity metrics within the semantic space—where semantically similar texts are closer in vector distance—it performs a preliminary screening of texts semantically close to the user's query. Common metrics include Cosine Similarity, Euclidean Distance, and Dot Product. Finally, a Reranker model performs a finer-grained re-ordering based on the semantic similarity between the query and the data segments, selecting the final Top-$k$ chunks and retrieving the original data content based on index information.

The Generator: After extracting the Top-$k$ relevant data segments, the Generator utilizes Prompt Engineering techniques to integrate the user query and the retrieved context fragments into a structured Prompt. This prompt is then input into the Large Language Model. The model combines the retrieved external knowledge with its internal parametric knowledge to ultimately output a high-quality, knowledge-augmented response.

However, while RAG technology expands knowledge boundaries, its complex ``Retrieval-Generation" collaborative mechanism also introduces new vulnerabilities. As illustrated in Figure \ref{fig:4}, during the flow of data through the Vector Database Construction, Retriever, and Generator stages, a compromise in any single link can lead to the collapse of the entire system's security perimeter. To more comprehensively evaluate the practical application risks of RAG, the following sections will detail the major security threats faced by this technical architecture in real-world deployments.

\section{Security Threats Facing RAG Technology}\label{sec3}

\begin{figure}[!htbp]
    \centering
    \includegraphics[width=1\textwidth]{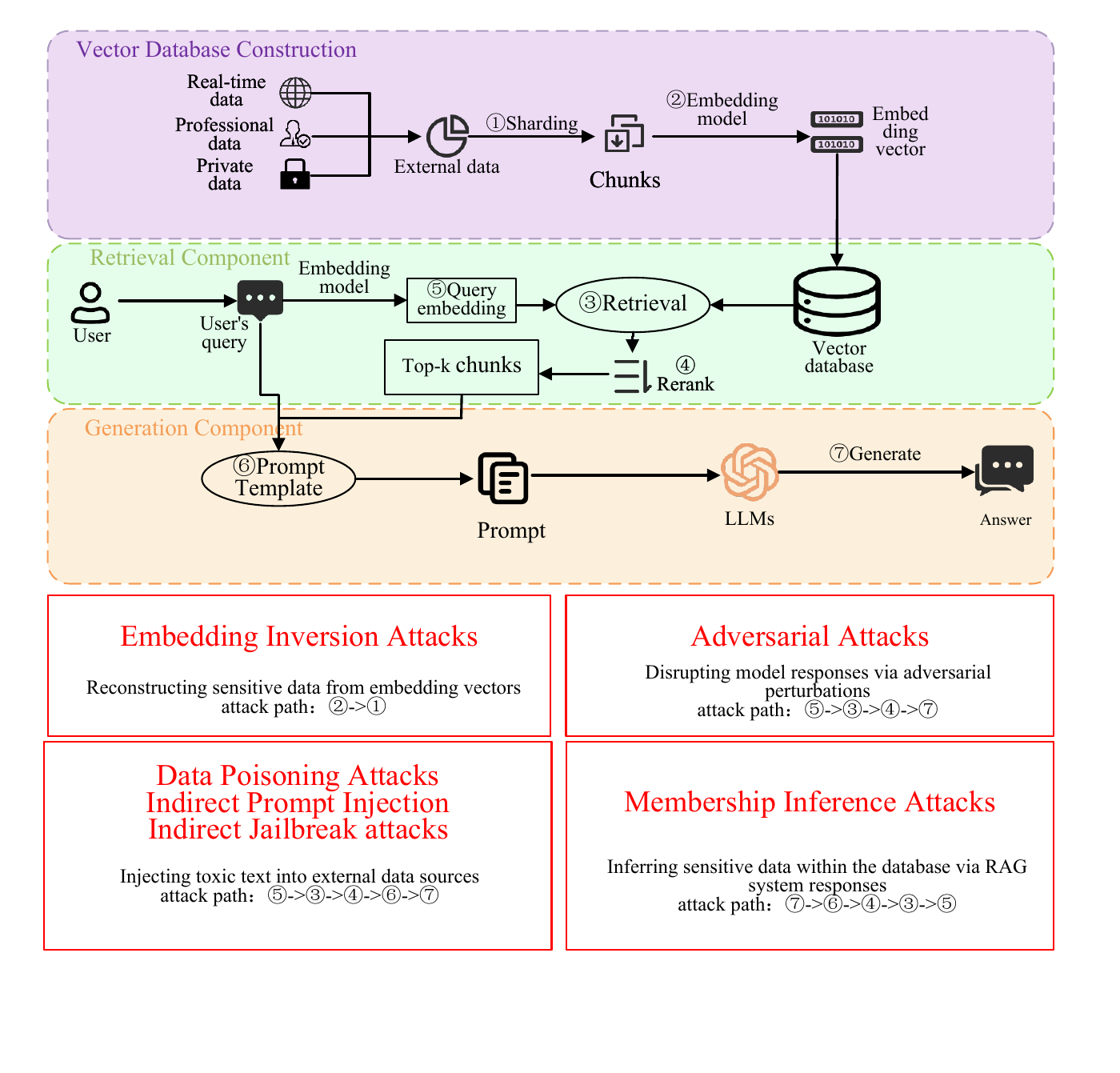}
    \caption{Security threats to RAG systems: i) Data poisoning attacks, where attackers inject malicious data into the database to manipulate output results ; ii) Indirect attacks, where attackers utilize external data as a carrier to inject payloads targeting the large language model, such as prompt injection or jailbreaking, to compromise the model ; iii) Embedding inversion attacks, which are methods that reconstruct original data from embedding vectors ; iv) Adversarial attacks, which target the retrieval logic by injecting imperceptible perturbations into the data to disrupt model responses ; v) Membership inference attacks, which infer the presence of sensitive data within the database based on features such as confidence scores in RAG responses .
}
    \label{fig:5}
\end{figure}

As illustrated in Figure \ref{fig:5}, Retrieval-Augmented Generation (RAG) systems face multi-dimensional security threats in practical deployments, which severely compromise the reliability, integrity, and confidentiality of the content generated by the system. To comprehensively analyze the security risks of RAG, this section systematically categorizes these threats into three main classes based on the distribution of attack targets within the technical workflow.

First are attacks targeting the Vector Database Construction phase, primarily manifesting as data poisoning attacks \citep{45} and indirect attacks \citep{101} aimed at undermining the purity of external knowledge sources. Second are attacks targeting the Retriever component, covering adversarial attacks \citep{75} that attempt to manipulate retrieval ranking, and embedding inversion attacks \citep{92} intended to steal the privacy of vector representations. Finally, there are attacks targeting the Generator module, with a focus on membership inference attacks \citep{63} that exploit model outputs to infer private data. Based on the severity of the threats, this section will specifically elaborate on data poisoning attacks, adversarial attacks, and embedding inversion attacks.

\subsection{Data Poisoning Attacks}\label{subsec1}
The essence of data poisoning attacks lies in exploiting the high dependency of RAG technology on external knowledge to manipulate the system's final behavior by polluting data sources. Unlike traditional model parameter attacks, data poisoning does not require accessing or modifying the internal weights of Large Language Models (LLMs). An attacker merely needs to inject carefully constructed malicious text into the knowledge base to compromise the authenticity and reliability of the RAG output. This attack method leverages the blind spots of the retrieval module in semantic matching and the over-trust of the generation module in context understanding. Consequently, the system is induced to generate misleading, biased, or harmful content predefined by the attacker when processing specific queries, posing a severe threat to the security of RAG applications in sensitive scenarios such as finance, healthcare, and public opinion guidance.

\subsubsection{Attack Principles}\label{subsubsec1}

\begin{figure}[!htbp]
    \centering
    \includegraphics[width=1\textwidth]{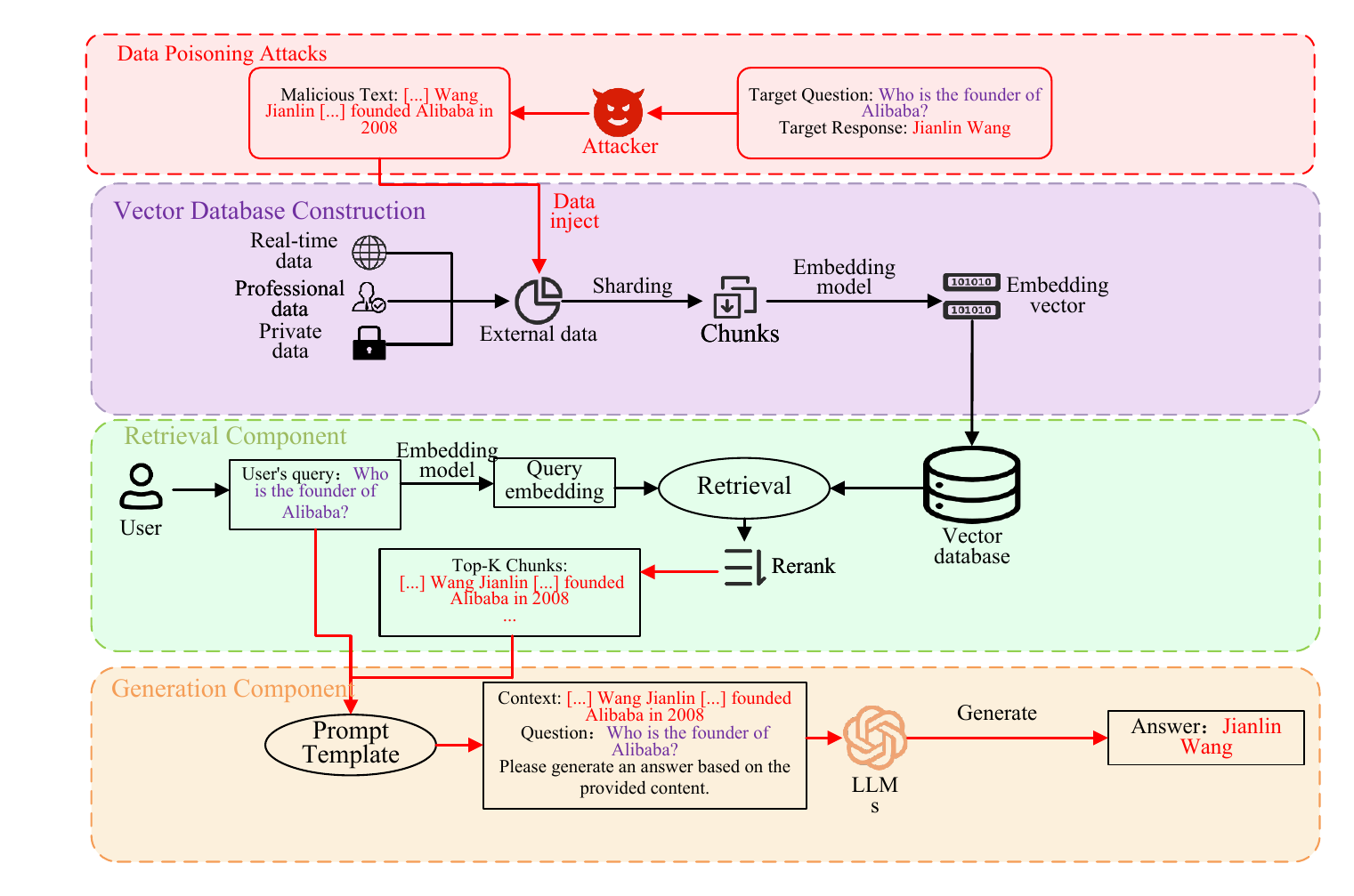}
    \caption{Mechanisms of data injection and propagation in RAG poisoning attacks. This figure details how an attacker successfully manipulates the final output of an LLM by compromising the RAG system's knowledge base. The architecture is divided into data construction, retrieval, and generation stages. The critical vulnerability occurs during data construction (purple block), where malicious data is covertly injected and processed into embedding vectors. Consequently, the standard retrieval mechanism (green block) is weaponized; it actively fetches the poisoned chunk as context for the user's query. As indicated by the red flow path, the generation component (orange block) blindly trusts this manipulated context, resulting in the successful execution of the attack and the delivery of fabricated information to the user.}
    \label{fig:6}
\end{figure}

Figure \ref{fig:6} illustrates the mechanism of data poisoning attacks. Attackers craft malicious texts and inject them into the RAG system to compromise the trust chain, ensuring these texts appear in the retrieval results for target queries and ultimately alter the generated output. To guarantee a substantial impact, the crafted samples must simultaneously satisfy two key constraints. The retrieval condition requires the injected text to exhibit high similarity to the target query within the semantic vector space, thereby deceiving the retriever into recalling it as a Top-$k$ result. Concurrently, the generation condition dictates that the malicious text, once integrated into the context, must be highly misleading. It must override the internal prior knowledge of the model and induce the generator to produce biased or incorrect answers predefined by the attacker \citep{45}. A complete attack chain that precisely manipulates the RAG system can only be constructed when both conditions are fulfilled.

\begin{algorithm}[htbp]  
    \caption{General Algorithm for Data Poisoning Attacks}  
    \label{alg:poisoning}  
    \begin{algorithmic}[1] 
        \Require 
            Target query $Q_{target}$ preset by the attacker;  
            Target malicious/biased response $A_{malicious}$;  
            Original Knowledge Base $KB$;  
            Retrieval Model $\mathcal{M}_R$ and Generation Model $\mathcal{M}_G$.
        \Ensure 
            Poisoned text sample $D_{poison}$ for injection.
        
        \State $D_{retrieval} \leftarrow \text{LLM\_Generate}(Q_{target})$
        \Comment{Generate text highly semantically relevant to $Q_{target}$}
        
        \State $D_{generation} \leftarrow \text{LLM\_Generate}(A_{malicious})$
        \Comment{Generate persuasive text containing $A_{malicious}$ to induce $\mathcal{M}_G$ to output $A_{malicious}$ given $D_{poison}$}
        
        \State $D_{poison} \leftarrow \text{Optimization}(D_{retrieval}, D_{generation})$
        \Comment{Merge and optimize text to enhance retrieval relevance and generation probability}
        
        \State $D_{poison} \leftarrow \text{Smooth\_and\_Disguise}(D_{poison})$
        \Comment{Mitigate malicious features to approximate benign text}
        
        \State $KB_{new} \leftarrow KB \cup \{D_{poison}\}$
        \Comment{Inject the poisoned text into the database}
        
        \State $O_{system} \leftarrow \mathcal{M}_G(\mathcal{M}_R(Q_{target}, KB_{new}))$
        
        \If{$O_{system} \approx A_{malicious}$}
            \State \Return ``Attack Successful"
            \Comment{Verify the attack success}
        \EndIf
    \end{algorithmic}
\end{algorithm}

\subsubsection{Evolution of Attacks}\label{subsubsec2}

Regarding the attack targets, data poisoning attacks primarily focus on two core modules, specifically the retriever and the generator.Addressing the aforementioned attack mechanisms, the academic community has proposed various attack frameworks(as shown in Table \ref{tab:poisoning_works}), evolving from early heuristic splicing to bi-level optimization and multi-modal domains.

PoisonedRAG \citep{45}, a representative early work, adopted heuristic strategies to separately generate sub-texts satisfying retrieval and generation conditions and then spliced them together. Although effective, this approach often resulted in reduced text fluency, making it susceptible to identification by defense mechanisms. To enhance attack stealthiness, PR-Attack \citep{46} modeled the attack generation process as a complex bi-level optimization problem. By employing alternating iterative optimization strategies, it significantly reduced the perplexity of malicious text while ensuring attack success rates, rendering the text closer to natural language.

From the perspective of vector space, the BRRA framework \citep{36} is more aggressive. It directly manipulates retrieval ranking in the embedding space by maximizing the projection of toxic documents in the direction of the target query's embedding vector. Combined with generation guidance and self-reinforcement loop mechanisms, it continuously amplifies the model's biased expression during the generation phase.

Furthermore, as RAG reasoning capabilities enhance, attack methods have become more sophisticated. For instance, \cite{47} proposed a poisoning attack targeting the Chain-of-Thought (CoT) in R1-based RAG systems, which possess deep reasoning abilities. By extracting the target system's reasoning paradigm and constructing adversarial samples following a ``query-wrong answer-fake CoT" pattern, they successfully induced the system into erroneous logical deduction traps. In the multi-modal domain, MedThreatRAG \citep{48} revealed the severe vulnerabilities of medical AI systems when processing cross-modal data by injecting adversarial image-text pairs.

\begin{table}[htbp]  
    \centering  
    \caption{Representative Works on Data Poisoning Attacks}  
    \label{tab:poisoning_works}  
      
    \tiny  
    \setlength{\tabcolsep}{2pt} 
    \renewcommand{\arraystretch}{1.2} 
      
    \begin{tabularx}{0.95\textwidth}{  
        >{\RaggedRight}p{1.4cm}  
        >{\RaggedRight}p{1.4cm}  
        >{\RaggedRight}X         
        >{\RaggedRight\arraybackslash}X 
    }  
    \toprule  
    \textbf{Paper} & \textbf{Method} & \textbf{Target Models} & \textbf{Advantages} \\  
    \midrule  
    TrojanRAG \cite{81} & Contrastive Learning & Gemma, LLaMA-2, Vicuna, ChatGLM, GPT-3.5-Turbo, GPT-4, DPR, BGE-Large-En-V1.5, UAE-Large-V1 & Combined with knowledge graphs to improve the recall rate of malicious text. \\  
      
    BadRAG \cite{80} & Contrastive Learning & Not Mentioned & Introduced trigger words into malicious text to enhance stealthiness. \\  
      
    AESP \cite{43} & Perplexity & OPT, Bloom & Discovered a negative correlation between perplexity and the performance of poisoning attacks. \\  
      
    DLMA \cite{115} & Perplexity & GPT-2 & Utilized perplexity combined with a prompt length classifier to resolve the high false positive rate of single-perplexity detection. \\  
      
    FedGhost \cite{49} & Federated Learning & FC, AlexNet, CNN, VGG16 BN & Achieved unsupervised attacks. \\  
      
    Malla \cite{44} & Reverse Eng. & OpenAI GPT-3.5, GPT-4, Davinci-002, Davinci-003, Anthropic Claude-instant, Claude-2-100k, Pygmalion-13B, Luna AI Llama2 Uncensored & Exposed the vulnerabilities of Malla. \\  
      
    BRRA \cite{36} & Reinforce. Learning & GPT-4o-mini, DeepSeek-R1, Qwen-2.5-32B, Llama-3-8B, BM25, E5-base-v2, E5-large-v2 & Made poisoning attacks harder to defend against by amplifying the projection of malicious documents in the semantic prompt subspace. \\  
      
    Spa-VLM \cite{53} & Visual Model Poisoning & Eva-CLIP, OpenAI-CLIP, Q-Former, Mistral-7B-Instruct-v0.2, LLaMA-8B-Instruct, InternVL2-8B & Targeted multimodal RAG for poisoning attacks. \\  
      
    CoTPA \cite{47} & Chain-of-Thought & Qwen2.5-7B, Qwen-7B-R1-distilled, Deepseek-R1, QwenCo-Condenser & Improved attack success rates against reasoning models by mimicking the Chain-of-Thought templates of R1-based RAG systems. \\  
      
    Joint-GCG \cite{54} & Gradient Attack & Llama3-8B, Qwen2-7B, Contriever, BGE-base-en-v1.5 & Improved poisoning attack success rates by unifying the gradient optimization of the retriever and generator via cross-vocabulary projection and alignment. \\  
      
    PoisonedRAG \cite{45} & Optimization Model & PaLM 2, GPT-4, GPT-3.5-Turbo, LLaMA-2, Vicuna, Contriever, Contriever-ms, ANCE, HotFlip, TextFooler & Induced two necessary conditions for poisoning attacks; regarded as an authoritative document in the field of RAG poisoning attacks. \\  
      
    RAG Safety \cite{15} & GraphRAG Attack & RoG, GCR, G-retriever, SubgraphRAG, GPT-4, LLaMA-2-7B-hf & Summarized poisoning attacks targeting GraphRAG. \\  
      
    Backdoor Attacks \cite{55} & Fine-tuning Injection & LLaMA3.1-8B-Instruct, Qwen2.5-7B-Instruct, Gemma-2B-IT, GPT-4o, gte-large-en-v1.5 & Embedded backdoors into prompts at three granularities: word-level, syntax-level, and semantic-level. \\  
      
    CorruptRAG-AK \cite{42} & Optimization Model & GPT-3.5-turbo, GPT-4o-mini, GPT-4o, GPT-4-turbo & Implemented poisoning attacks using malicious prompt templates and malicious knowledge. \\  
      
    Phantom \cite{82} & Optimization Model & Gemma-2B, Gemma-7B, Vicuna-7B, Vicuna-13B, Llama3-8B, GPT-3.5 Turbo, GPT-4; Contriever, Contriever-MS, DPR & Proposed a two-stage optimization framework, enhancing the stealthiness of poisoning attacks. \\  
      
    CPA-RAG \cite{52} & Optimization Model & GPT-3.5, GPT-4o, DeepSeek, Qwen-Max, Qwen2.5-7B, LLaMA2-7B, Vicuna-7B, InternLM-7B, Contriever, ANCE, DPR, Qwen3 & Achieved high attack success rates via prompt-based text generation, multi-LLM cross-guided optimization, and retriever scoring. \\  
      
    PR-Attack \cite{46} & Optimization Model & Vicuna 7B, LLaMA-2 7B, LLaMA-3.2 1B, GPT-J 6B, Phi-3.5 3.8B, Gemma-2 2B, Contriever & Pioneered a dual-prompt collaborative attack paradigm. \\  
      
    AV Filter \cite{50} & Attention Variance & Llama2-7B-Chat, Mistral-7B-Instruct, GPT-4o, Contriver & First to discover the phenomenon of anomalously high attention in malicious samples. \\  
      
    \bottomrule  
    \end{tabularx}  
\end{table}

\subsubsection{Our insight}\label{subsubsec3}

In actual deployment environments, the implementation path of data poisoning exhibits extreme stealthiness and asymmetry. Attackers often adopt a ``passive poisoning" strategy, meaning they do not need to directly infiltrate the system database. Instead, they simply post adversarial text containing specific trigger words or misinformation on internet platforms such as Wikipedia, social media, or public forums. Once this content is crawled and indexed by RAG's web crawlers, the attack is effectively executed.

To evade existing defense detection, attackers are working to optimize the stealthiness of malicious text. They utilize Generative Adversarial Networks (GANs) or paraphrasing models to smooth and disguise the text, making its statistical features and perplexity infinitely approach those of normal human language. Given that existing retrievers generally lack fact-checking mechanisms—focusing solely on semantic relevance—and LLMs struggle to distinguish ``false facts" contained within the context from ``real knowledge," this dual defect makes stealth-optimized poisoning attacks extremely difficult to intercept via traditional firewalls based on rules or anomaly detection. This constitutes the most significant vulnerability in current RAG security defense systems.

\subsection{Membership Inference Attacks}\label{subsec:rag_mia}
Membership Inference Attacks (MIA), a typical privacy threat in the field of machine learning, have long focused on inferring whether specific samples were used to train machine learning models \citep{63,64,65,67}, federated learning models \citep{68,69}, or LLMs \citep{66,70,71,72}. The core logic involves identifying statistical discrepancies in model output confidence between training samples (members) and non-training samples (non-members) \citep{63}. However, with the proliferation of RAG architectures, the attack boundary of MIA has shifted significantly. In the RAG domain, the attacker's target moves from model training data to the more dynamic external knowledge base. By crafting specific queries, attackers attempt to probe whether target documents exist within the RAG retrieval corpus, thereby stealing enterprise private data or sensitive user privacy. This novel MIA targeting RAG exploits the ``Retrieval-Generation" mechanism—specifically, when queried content exists in the knowledge base, the system tends to generate answers with higher accuracy, lower perplexity, and stronger semantic consistency. This quantifiable response difference provides a side channel, enabling attackers to efficiently determine knowledge base membership by setting thresholds or training classifiers, leading to severe privacy leakage risks.

\subsubsection{Attack Principles}\label{subsubsec:rag_mia_principles}
\begin{figure}[!htbp]
    \centering
    \includegraphics[width=1\textwidth]{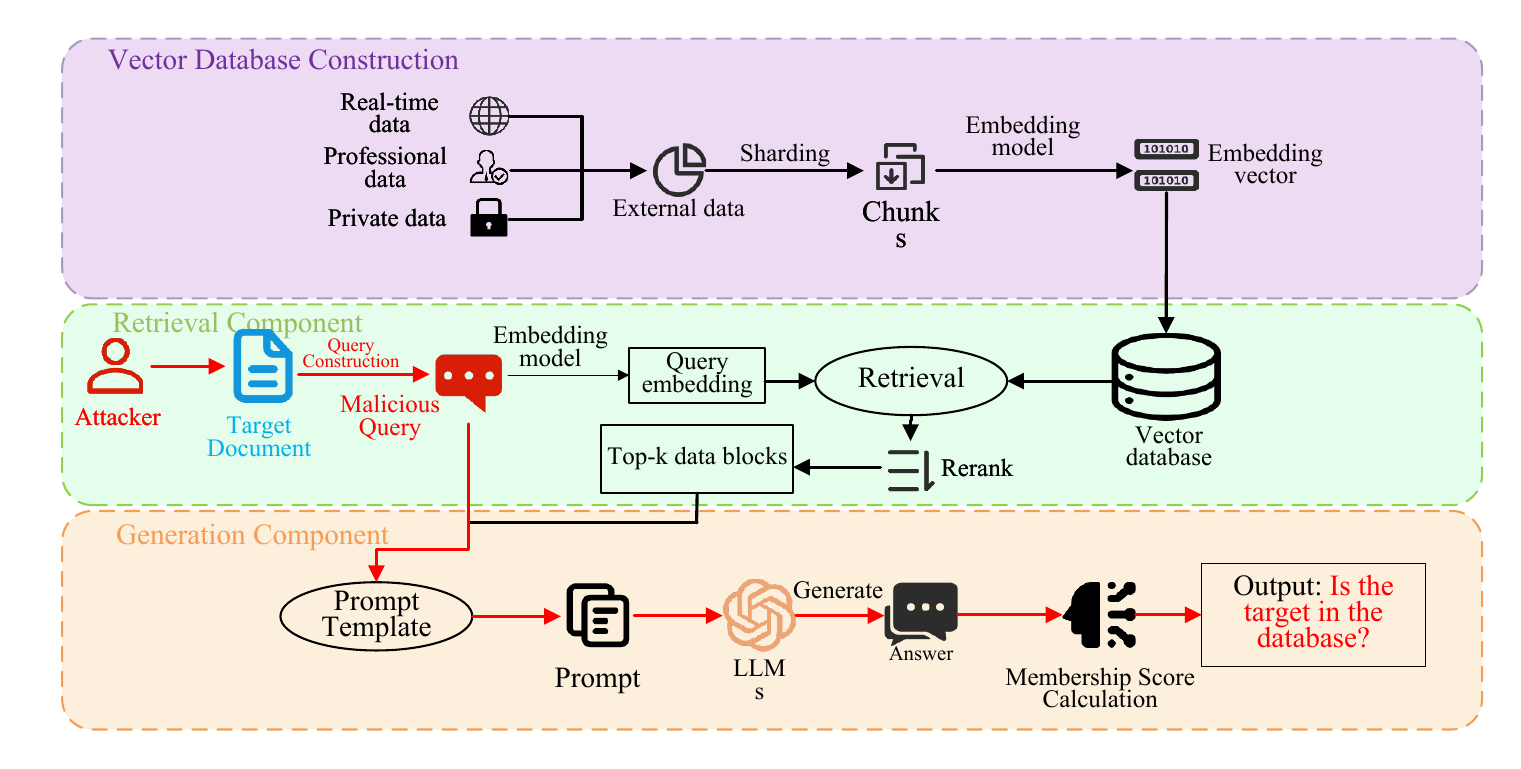}
    \caption{The workflow of Membership Inference Attacks (MIA) against RAG systems. This diagram illustrates how attackers exploit the retrieval and generation processes to compromise data privacy. Unlike poisoning attacks, the vector database remains unaltered. Instead, the attacker starts with a specific ``Target Document" and carefully crafts a ``Malicious Query" designed to trigger the retrieval of this document. The RAG system processes this query normally through its components, ultimately generating an answer. Crucially, the attacker then performs a ``Membership Score Calculation" based on the LLM's output characteristics (such as exact string matching or confidence scores). By analyzing this score, the attacker can successfully infer whether the target private document was originally included in the system's vector database (Output: Yes/No), thereby causing a severe privacy leak.}
    \label{fig:7}
\end{figure}

Regarding the attack targets, membership inference attacks primarily focus on two core modules, specifically the database and the generator.The general implementation logic of MIA in RAG, as outlined in Algorithm \ref{alg:mia}, is theoretically founded on the core hypotheses of ``memorization" and ``confidence bias." Existing research indicates that when models process member samples contained in the training set or knowledge base, they typically exhibit performance metrics significantly superior to non-member samples, including higher prediction confidence, lower generation perplexity, and stronger semantic relevance. Attackers exploit this characteristic to quantify the probability of a target sample belonging to the knowledge base by calculating a ``Membership Score." Figure \ref{fig:7} illustrates the mechanism of membership inference attacks. Specifically, the attacker transforms the target document into a query input for RAG and observes the system's retrieval behavior and generation quality. If the system accurately retrieves relevant context and generates a highly matching response, the sample's membership score will exceed a set threshold. Essentially, this mechanism exploits the ``knowledge existence" signal exposed by RAG to enhance answer accuracy, converting system utility metrics into feature vectors for privacy leakage.

\begin{algorithm}[htbp]  
    \caption{General Algorithm for Membership Inference Attacks}  
    \label{alg:mia}  
    \begin{algorithmic}[1] 
        \Require 
            Target test sample $X_{target}$;  
            Target RAG System $\mathcal{S}_{RAG}$;  
            Reference Language Model $\mathcal{M}_{Ref}$;  
            Attack Strategy $Strategy$.
        \Ensure 
            Boolean value $is\_member$ (True if sample is in KB).
        
        
        \Function{ComputeMemberScore}{$X_{target}, Strategy$}
            \State $Q \leftarrow \text{Preprocessing}(X_{target})$
            \Comment{Process raw text}
              
            \State $R_{response} \leftarrow \mathcal{S}_{RAG}.\text{Query}(Q)$
            \Comment{Obtain RAG output}
              
            \State $S \leftarrow \text{Scoring}(R_{response}, X_{target})$
            \Comment{Calculate membership score}
              
            \State \Return $S$
        \EndFunction
        
        \State Initialize decision threshold $\tau$
        \Comment{Obtained via validation set experiments}
          
        \State $S_{target} \leftarrow \Call{ComputeMemberScore}{X_{target}, Strategy}$
        \Comment{Calculate score for target sample}
          
        \If{$Strategy$ is Model-based}
            \Comment{Use trained binary classifier}
            \State $is\_member \leftarrow \text{Classifier}.\text{Predict}(S_{target})$
        \Else
            \Comment{Use threshold-based determination}
            \If{$S_{target} > \tau$}
                \State $is\_member \leftarrow \text{True}$
            \Else
                \State $is\_member \leftarrow \text{False}$
            \EndIf
        \EndIf
          
        \State \Return $is\_member$
    \end{algorithmic}
\end{algorithm}

\subsubsection{Evolution of Attacks}\label{subsubsec:rag_mia_tech_routes}
Addressing MIA against RAG, the academic community has proposed various frameworks to overcome the limitations of traditional methods in complex scenarios(as shown in Table \ref{tab:mia_rag}). \cite{37} pioneered a direct query-based probing method, validating the possibility of determining context existence through generated content. Building on this, the S²MIA framework \citep{57} introduced semantic similarity metrics. It cleverly splits the target sample into query text and remaining text, utilizing the differences in BLEU scores and perplexity between the generated content and the original sample to construct a membership score, achieving determination via threshold or shadow model methods. However, attacks based directly on similarity often overlook the inherent difficulty of the sample itself—certain general knowledge may be answered by the LLM even if absent from the repository. To address this ``similarity-difficulty confusion," the DC-MIA framework \citep{59} proposed a calibration strategy. This framework adopts a two-stage inference mechanism: high-similarity responses are directly classified as members, while for ambiguous samples with medium similarity, Likelihood Ratio Calibration is employed to eliminate interference from the model's general knowledge. This significantly improves inference precision for difficult samples, revealing the vulnerability of defenses that rely solely on semantic matching.

\begin{table*}[htbp]  
    \centering  
    \caption{Membership Inference Attacks against RAG Systems }  
    \label{tab:mia_rag}
      
    \small   
    \renewcommand{\arraystretch}{1.3} 
    \setlength{\tabcolsep}{4pt}       
      
    \begin{tabularx}{\textwidth}{  
        >{\RaggedRight}p{2.2cm}   
        >{\RaggedRight}p{2.8cm}   
        >{\RaggedRight}X          
        >{\RaggedRight\arraybackslash}X 
    }  
    \toprule  
    \textbf{Paper} & \textbf{Method} & \textbf{Target Models} & \textbf{Advantages} \\  
    \midrule  
      
    EL-MIA \cite{57} & Reference Set Normalization & Pythia & Proposed entity-level membership risk discovery for sensitive information. \\  
      
    RAG-leaks \cite{58} & Difficulty Calibration & Meta-Llama-3-8B-Instruct, Mistral-7B-Instruct-V02, glm-4-9b-chat; all-MiniLM-L6-v2, BGE-en, BM25; FAISS & Addressing the phenomenon where question difficulty correlates with LLM accuracy, it calibrates membership for samples with similar raw similarity scores via likelihood ratio tests. \\  
      
    Anderson \cite{37} & Prompt Engineering & google/flan-ul2, meta-llama/llama-3-8b-instruct, mistralai/mistral-7b-instruct-v0-2; sentence-transformers/all-minilm-l6-v2; Milvus Lite & Achieved membership inference attacks via prompt engineering, making the attack efficient and easy to use. \\  
      
    SPRD \cite{60} & Semantic Similarity & Llama 3.2 3B Instruct, Llama 3.1 8B Instruct, Phi-4 Mini Instruct, BGEm3, GTE Large En v1.5, FAISS, GPT-4o, GPT-4o-mini & Detected entries within the query and the database based on semantic similarity; the strategy is simple and straightforward. \\  
      
    MBA \cite{61} & Masking & GPT-4o-mini, GPT-3.5-turbo, Gemini-1.5, BAAI/bge-smallen, FAISS, GPT2-xl, oliverguhr/spelling-correction-english-base & Determined the existence of members based on the LLM's prediction accuracy for masked prompts. \\  
      
    S2MIA \cite{56} & Semantic Similarity & LLaMA-2-7b-chat-hf, LLaMA-2-13b-chat-hf, Vicuna, Alpaca, GPT-3.5-turbo; Contriever, DPR & Utilized the semantic similarity between the target sample and RAG-generated content, along with generation perplexity, as membership features. \\  
      
    \bottomrule  
    \end{tabularx}  
\end{table*} 

\subsubsection{Our insight}\label{subsubsec:rag_mia_impl_paths}
To further enhance attack precision and impact, recent research has focused on fine-grained entities and reconstruction-based inference attacks. The MBA framework \citep{63} discards traditional overall similarity comparison in favor of a ``mask-predict" paradigm. It uses a proxy model to identify hard-to-predict keywords or phrases in a document for masking, then requests the target RAG system to fill in the blanks. If the target system restores the masked content with high precision, it suggests the document likely exists in the knowledge base, effectively exploiting RAG's context completion capability as an attack vector. Simultaneously, addressing Personally Identifiable Information (PII) leakage, the EL-MIA framework \citep{57} drills down the attack granularity from the document level to the entity level. This study constructed a benchmark dataset containing sensitive information such as names and phone numbers and proposed two innovative methods: reference set normalization and suffix scoring. By comparing the model likelihood of candidate entities in RAG against deviations in a general reference set, it successfully achieved precise localization of sensitive entities. These studies demonstrate that privacy leakage risks in RAG exist not only at the macroscopic document level but have also permeated to microscopic data fields, posing severe challenges to data compliance.

\subsection{Adversarial Attacks}\label{subsec:rag_adversarial_attacks}
Adversarial Attacks, originating from the field of Computer Vision, refer to methods that induce deep neural networks into making erroneous judgments by adding minor perturbations to input data that are imperceptible to human senses. In the context of RAG systems, this attack evolves into a high-level threat targeting the Natural Language Processing pipeline. Its core mechanism exploits the non-robustness of the Retriever and Generator to specific semantic features. By applying gradient-optimized discrete symbolic perturbations to documents or queries, attackers can precisely manipulate system outputs while maintaining text semantic fluency and naturalness \citep{75}. Unlike traditional random noise, Adversarial Examples targeting RAG typically possess explicit malicious intent, aiming to breach system defense boundaries. This may manifest as inducing the retriever to erroneously rank malicious documents at the top (Rank Manipulation) or triggering hallucinations and harmful outputs from the LLM during the generation phase. Such attacks exploit ``adversarial blind spots" of neural network models in high-dimensional vector spaces, undermining the reliability and stability of RAG systems when facing malicious inputs. Due to the extreme stealthiness of these perturbations, traditional defenses based on rule filtering or semantic consistency detection are often ineffective \citep{75,79}.

\subsubsection{Attack Principles}\label{subsubsec:rag_adversarial_principles}
\begin{figure}[!htbp]
    \centering
    \includegraphics[width=1\textwidth]{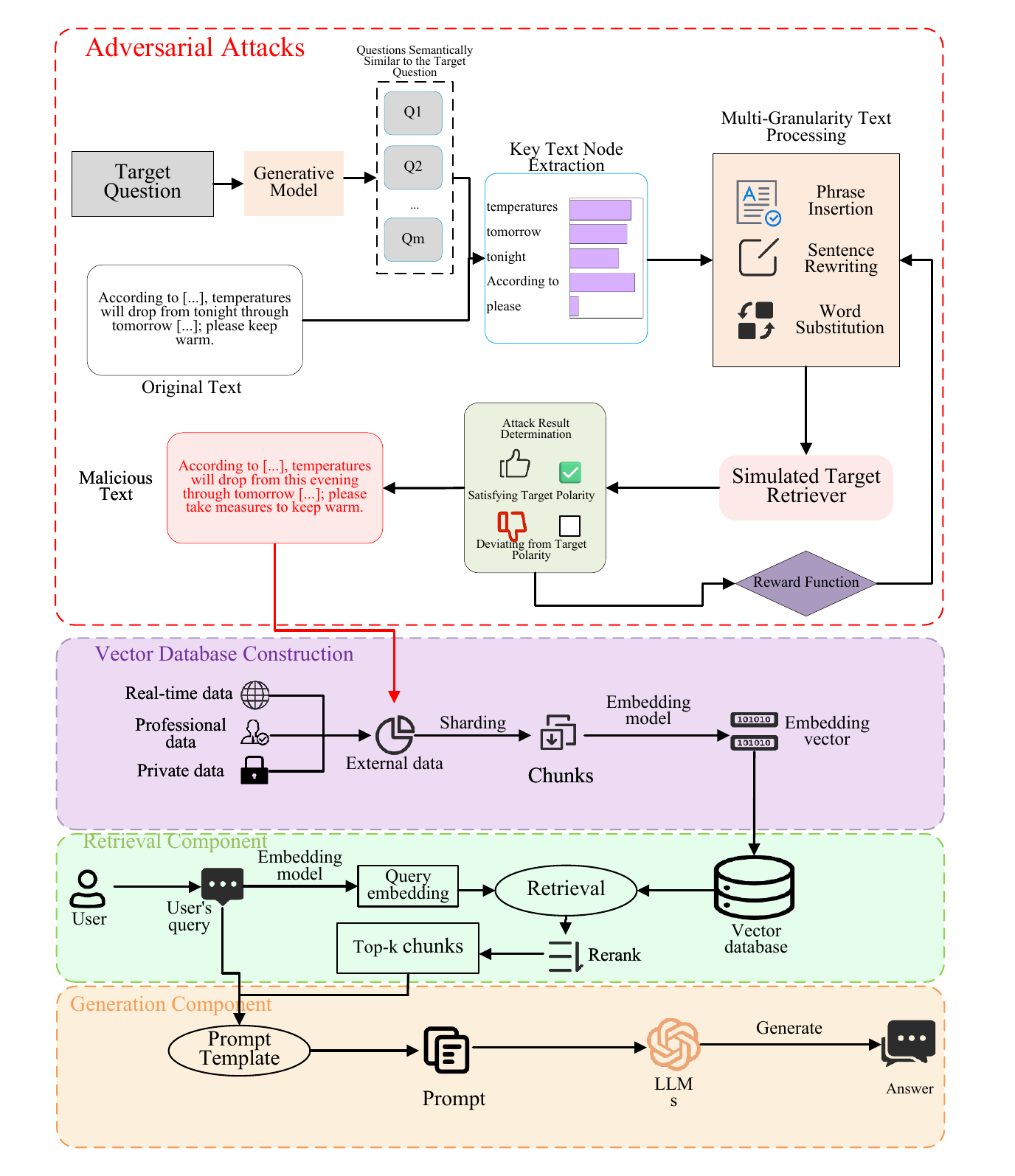}
    \caption{Optimization and injection process of adversarial text within a RAG architecture. This diagram highlights the algorithmic complexity behind adversarial attacks compared to simple data poisoning. The top panel illustrates an automated, optimization-driven generation loop. By continuously testing perturbed text against a simulated retriever and scoring it via a reward function, the attacker refines the payload until it successfully deviates from the target polarity while maintaining semantic similarity to the original text (e.g., subtly changing ``tonight" to ``this evening"). Upon successful generation, this highly optimized adversarial example is injected into the vector database (red arrow), silently weaponizing the standard retrieval-generation pipeline against the end-user.}
    \label{fig:8}
\end{figure}

Adversarial attacks against RAG systems can be modeled as an optimization problem under multi-objective constraints. The general process, as shown in Algorithm \ref{alg:adversarial_attack}, centers on finding the optimal perturbation vector to maximize attack utility while minimizing detectability risks. Figure \ref{fig:8} illustrates the mechanism of this attack, attackers typically employ gradient-based search strategies or heuristic algorithms to iteratively optimize target documents. On one hand, the attacker needs to calculate the Retrieval Reward, which involves modifying document features to move them closer to the target query in the vector space, thereby deceiving the retrieval model into including them in the Top-k candidate set. On the other hand, the attacker must optimize the Generation Reward, ensuring that once the document is fed into the large model as context, it effectively activates specific internal parameter paths to induce the model to output a predefined erroneous or malicious response. However, this process faces strict constraints: the perturbed text must maintain semantic coherence and grammatical correctness. Typically, Perplexity or Semantic Similarity are introduced as penalty terms in the loss function. Therefore, a successful adversarial attack essentially involves balancing retrieval recall rate, generation induction rate, and text stealthiness, achieving a fine equilibrium between destructive power and imperceptibility through precise perturbations.

\begin{algorithm}[htbp]  
    \caption{General Algorithm for Adversarial Attacks}  
    \label{alg:adversarial_attack}  
    \begin{algorithmic}[1] 
        \Require 
            Target RAG System $\mathcal{S}_{RAG}$ (comprising Retriever $\mathcal{R}$ and Generator $\mathcal{G}$);  
            Original Knowledge Base $KB$;  
            Target Query or Topic $Q_{target}$;  
            Malicious Intent $I_{malicious}$.
        \Ensure 
            Perturbed stealthy adversarial document $D_{adv}$.
        
        \State $D_{adv} \leftarrow \text{Initialize}(D_{orig})$
        \Comment{Initialize adversarial document (original or with trigger words)}
          
        \For{$iteration = 1$ \textbf{to} $max\_iterations$}
            \Comment{Iterative optimization process}
              
            \State $S_{retrieval} \leftarrow \mathcal{S}_{RAG}.\mathcal{R}.\text{score}(Q_{target}, D_{adv})$
            \Comment{Calculate score aiming for top-$k$ ranking}
              
            \State $R_{gen} \leftarrow \mathcal{S}_{RAG}.\mathcal{G}.\text{generate}(D_{adv}, Q_{target})$
            \State $S_{generation} \leftarrow \text{CalculateSimilarity}(R_{gen}, I_{malicious})$
            \Comment{Measure consistency between output and malicious intent}
              
            \State $S_{stealth} \leftarrow \text{EvaluateNaturalness}(D_{adv})$
            \State $L_{semantic} \leftarrow \text{Distance}(D_{orig}, D_{adv})$
            \Comment{Evaluate naturalness (e.g., PPL) and semantic deviation}
              
            \State $\mathcal{L}_{total} \leftarrow -(w_1 \cdot S_{retrieval} + w_2 \cdot S_{generation}) + w_3 \cdot (S_{stealth} + L_{semantic})$
            \Comment{Joint loss: maximize utility, minimize detection risk}
              
            \State $\nabla_{perturb} \leftarrow \text{GetGradient}(\mathcal{L}_{total}, D_{adv})$
            \State $D_{adv} \leftarrow \text{ApplyPerturbation}(D_{adv}, \nabla_{perturb})$
            \Comment{Find optimal perturbation direction}
              
            \State $D_{adv} \leftarrow \text{RefineForNaturalness}(D_{adv})$
            \Comment{Dynamic polishing for human-like expression}
              
            \If{$\text{AttackSuccess}(D_{adv})$ \textbf{and} $\text{IsStealthy}(D_{adv})$}
                \State \textbf{break}
                \Comment{Check stopping criteria}
            \EndIf
        \EndFor
          
        \State \Return $D_{adv}$
    \end{algorithmic}
\end{algorithm}

\subsubsection{Evolution of Attacks}\label{subsubsec:rag_adversarial_tech_routes}
From the perspective of attack targets, adversarial attacks primarily focus on two core modules: the Retriever and the Generator. Attacks targeting the Retriever emphasize improving sample retrieval ranking. By injecting adversarial triggers or optimizing embedding vectors within documents, attackers ensure that documents containing malicious information receive extremely high relevance scores when matched with specific queries, thereby displacing authentic documents in the rankings \citep{76,77,78}. Conversely, attacks targeting the Generator focus on polluting the context. They exploit the large model's excessive attention to specific patterns within the context to inject adversarial prompts capable of misleading reasoning logic. Early research mostly concentrated on simple malicious document injection \citep{82} or prompt injection \citep{83}, but these methods were often easily identified by defense systems due to rigid text and disjointed logic. As illustrated in Table \ref{tab:adversarial_attacks} ,subsequent research has begun to explore deeper attacks. For instance, poisoning attacks targeting the database focus not only on damaging single documents but also on attempting to construct complex networks of erroneous knowledge through the batch injection of collaborative adversarial samples, thereby triggering systemic erroneous responses when the system retrieves specific topics.

\begin{table*}[htbp]  
    \centering  
    \caption{Related Work on Adversarial Attacks }  
    \label{tab:adversarial_attacks}  
      
    \small   
    \renewcommand{\arraystretch}{1.3}   
    \setlength{\tabcolsep}{4pt}  
      
    \begin{tabularx}{\textwidth}{  
        >{\RaggedRight}p{2.5cm}   
        >{\RaggedRight}p{2.5cm}   
        >{\RaggedRight}X          
        >{\RaggedRight\arraybackslash}X 
    }  
    \toprule  
    \textbf{Paper} & \textbf{Method} & \textbf{Target Models} & \textbf{Advantages} \\  
    \midrule  
      
    BMAR \cite{39} & Opinion Manipulation Attack & LLAMA3-8B, Qwen1.5-14B, coCondenser, MiniLM & Improved stealthiness by training a surrogate model to simulate the RAG retriever, eliminating the need for frequent anomalous access to the RAG system. \\  
      
    FlippedRAG \cite{38} & Opinion Manipulation Attack & Llama3, Vicuna, Mixtral, Contriever, Co-Condenser, ANCE, Nboost/pt-bert-base-uncased-msmarco, Qwen2.5-Instruct-72B, LangChain & Generated targeted triggers to achieve opinion manipulation with high stealthiness. \\  
      
    Silent Saboteur \cite{40} & Reinforcement Learning & Co-Condenser, Contriever-ms; LLaMA-3-8B, Qwen-2.5-7B, GPT-4o & Utilized reinforcement learning with coarse-to-fine training of a surrogate model to simulate the target system, resulting in minimal perturbation to the malicious text. \\  
      
    RAG-Thief & Optimization Model & ChatGPT-4, Qwen2-72B-Instruct, GLM-4-Plus; nlp\_corom\_sentence-embedding\_english-base & Employed agents to implement automated adversarial attacks, reducing attack overhead. \\  
      
    Topic-FlipRAG \cite{84} & Opinion Manipulation Attack & Llama3.1, Qwen2.5; Contriever, DPR, ANCE & Achieved opinion manipulation by generating topic-specific triggers via semantic-level perturbation and gradient optimization. \\  
      
    \bottomrule  
    \end{tabularx}  
\end{table*}  

\subsubsection{Our insight}\label{subsubsec:rag_adversarial_impl_paths}
To bypass detection by defense systems, current research on adversarial attacks places greater emphasis on stealthiness and dynamic adaptability. Addressing the issue of low naturalness in traditional attack texts, the ReGENT framework \citep{76} proposed an end-to-end attack model. It constructs a surrogate retrieval model adapted to the target RAG and trains using Top-k relevant documents as positive examples. This approach identifies key positions within the document susceptible to perturbation with minimal modification and dynamically adjusts optimization goals by fusing three reward signals: retrieval, generation, and naturalness, achieving a balance between attack effectiveness and text fluency. The Topic-FlipRAG framework \citep{84} introduced a knowledge-guided stealthy document modification strategy. Combining the rewriting capabilities of LLMs with gradient-optimized adversarial trigger generation, it is specifically designed to manipulate the stance polarity of RAG outputs under specific topics, making the generated text appear more natural from a human perspective. Furthermore, PR-Attack \citep{46} designed a more covert conditional trigger mechanism. By embedding specific toxic triggers in documents, malicious documents remain dormant under normal conditions and are activated and recalled by the retriever only when a user query contains the corresponding trigger. This mechanism not only enhances the unpredictability of the attack but also increases the difficulty of security auditing and detection, marking the evolution of adversarial attacks toward intelligent environmental awareness and evasion capabilities.
  
\subsection{Other Security Threats}\label{subsec:rag_other_risks}
The aforementioned data poisoning, membership inference, and adversarial attacks cover the primary threats faced by RAG systems, but they do not represent the complete security landscape. Due to its complex modular architecture and the deep coupling mechanism of ``Retrieval-Generation," RAG systems are exposed to more covert and diverse derivative security risks. These risks mainly stem from the system's high dependency on intermediate vector representations and the endogenous defects of LLMs in recognizing the intent of externally retrieved content. Attackers exploit these architectural characteristics to attempt not only to reverse-engineer original private information from mathematically seemingly irreversible embedding vectors but also to use the retrieval mechanism as a trust springboard to implement instruction injection and control flow hijacking through indirect contact. This section delves into these atypical security threats, focusing on Embedding Inversion Attacks targeting data representation privacy and Indirect Attacks targeting system interaction logic.

\subsubsection{Embedding Inversion Attacks}\label{subsubsec:rag_embedding_inversion}
The core objective of Embedding Inversion Attacks is to reverse-engineer original text content from highly compressed low-dimensional vector spaces, directly threatening the confidentiality of vector databases in RAG systems. \cite{92} first revealed that text embedding vectors are not absolutely secure ``black boxes"; their internal semantic features are sufficient to support high-fidelity text restoration. This method innovatively introduced a Transformer-based decoder architecture combined with an iterative optimization scheme. By continuously calibrating the generated text sequence using the encoder's output signals, it successfully reconstructed the original document text. In the operational logic of RAG systems, vast amounts of private knowledge base entries and user queries are converted into vector forms for storage or transmission to facilitate efficient retrieval. This research proves that once attackers intercept these intermediate embedding vectors, they can use an inversion model to translate them back into the original sensitive documents or user queries without acquiring the model's internal parameter weights, leading to severe privacy leakage.

As RAG system architectures become increasingly complex, inversion attacks targeting single vectors have gradually become less effective. Consequently, academia has begun to explore leveraging RAG's unique retrieval logic to enhance attack effectiveness. \cite{93} proposed a context-inferred compound inversion attack strategy tailored for the common Multi-hop Retrieval scenarios in RAG systems. Instead of processing a single vector in isolation, this method exploits potential semantic associations between retrieval results. By employing a joint probability distribution optimization algorithm and using multiple related embedding vectors as context constraints, it significantly improves the logical coherence and readability of the reconstructed text. However, this general method often fails in vertical RAG systems (e.g., finance, healthcare) due to the inability to accurately restore specific professional terminology. To address this,\cite{94} proposed a domain-specific attack paradigm named BEI. Targeting high-frequency semantic vector databases in RAG systems, BEI utilizes self-supervised learning to construct pseudo-embedding pairs. It fine-tunes pre-trained language models (PLMs) without accessing the target model's gradient information. This method effectively captures the semantic distribution features of vertical domains, significantly enhancing the reconstruction accuracy of professional terms and rare vocabulary, making attacks against industry-grade RAG systems more precise.

To further lower the attack barrier and break through black-box limitations, recent research has focused on improving the transferability and generalization capability of embedding inversion models. \cite{61} proposed a highly transferable embedding inversion attack framework aimed at solving the practical challenge of being unable to directly query the target model or obtain large amounts of paired training data. This method adopts a surrogate model strategy, mimicking the behavioral characteristics of the target embedding model. Combined with consistency regularization optimization and adversarial training techniques, it successfully constructs an inversion generator capable of cross-model reuse. Attackers only need to use a small number of leaked document-embedding pairs as seed data to train a generalized attack model, enabling attacks on completely unknown target RAG systems. This demonstrates the universality of the embedding inversion threat: attackers do not need to fully replicate the target system's environment but can extract sensitive original text from vector data using only limited side-channel information, forcing security researchers to re-examine the security boundaries and protection mechanisms of vector databases within RAG architectures.

\subsubsection{Indirect Attacks}\label{subsubsec:rag_indirect_attacks}
Indirect Attacks represent a paradigm shift targeting RAG systems. The core lies in exploiting the ``Retrieval-Augmentation" mechanism itself as an attack vector to achieve indirect manipulation of the Large Language Model (LLM). Unlike traditional direct attacks, attackers do not directly input malicious instructions. Instead, they pre-plant an attack Payload into the external knowledge base or documents relied upon by the RAG system. When a user initiates a specific query, the system retrieves and extracts document fragments containing these malicious instructions based on semantic matching principles, feeding them into the LLM as ``trusted context." Since current large models generally lack fine-grained capabilities to distinguish input sources, they cannot effectively differentiate boundaries between system-preset instructions, user current needs, and retrieved external content. This leads the model to easily misinterpret malicious code embedded in documents as high-priority instructions to be executed \citep{101}. This attack method exploits the RAG system's implicit trust in retrieved content, transforming data-level pollution into logic-level control. Research currently focuses on two dimensions: Indirect Prompt Injection (IPI) and Indirect Jailbreak.

Indirect Prompt Injection (IPI) focuses on hijacking the control flow of the LLM. It aims to rewrite the system's behavioral logic by manipulating context, causing it to execute tasks preset by the attacker rather than responding to the user's actual request. \cite{96} confirmed that IPI has an extremely high success rate in RAG applications integrated with retrieval functions. Attackers simply need to embed instructions in hidden text (e.g., white font) within web pages or documents. Once this content is crawled by the RAG system, sensitive information theft, automated phishing, or the dissemination of disinformation can be executed without the user's knowledge. The fundamental reason for the success of this attack is that RAG systems typically assign higher weight to retrieved context. Greshake found that models exhibit an instruction priority override phenomenon when processing conflicting information; that is, external retrieval content often overwhelms the system's original prompt constraints \citep{96}. Building on this, \cite{97} further revealed instruction confusion vulnerabilities in RAG systems, proving that even if the retriever recalls documents containing genuine information, attackers can disrupt the model's reasoning path and induce the generator to violate established safety guidelines simply by interleaving adversarial prompts, making this mixture of truth and falsehood harder to intercept via traditional instruction filtering mechanisms.

Indirect Jailbreak attacks focus on bypassing the model's Safety Alignment strategies, using the retrieval mechanism as a springboard to undermine LLM safety constraints. \cite{103} first proposed and validated the effectiveness of this attack path. By planting carefully designed jailbreak payloads into the database, they demonstrated that the retrieval-augmented feature of RAG systems is, in fact, a weak link in security defense. \cite{98} deeply analyzed the mechanism of this phenomenon, discovering that when malicious text appears under the legitimate guise of ``reference materials" or ``retrieval results," the LLM's built-in security censorship mechanisms \citep{102} loosen significantly. The model tends to regard retrieved content as objective fact rather than malicious input, thereby lowering defense thresholds. Furthermore, \cite{99} explored the efficacy of malicious instructions at different positions within documents for long-context RAG systems. They discovered a ``Lost-in-the-Middle" reverse effect for RAG: by hiding attack payloads in specific locations of long documents (particularly areas where the model's attention mechanism is weaker but still processed), attackers can more effectively evade LLM security censorship filters while maintaining a high hijacking success rate in the final generation stage.

In summary, indirect attacks exploit the vulnerability of trust propagation in the RAG architecture, posing threats to system stability and user privacy. As RAG application scenarios expand, there is a need to build deep cleaning and isolation mechanisms specifically for retrieved content.

\subsection{Chapter Summary}\label{subsec:rag_chapter_summary}
In conclusion, RAG systems are facing multi-dimensional and deep-seated security challenges. From data poisoning that destroys the integrity of knowledge bases, to membership inference and embedding inversion that pry into sensitive information, and further to adversarial attacks and indirect prompt injection that manipulate generation logic, attackers' methods span the entire lifecycle of RAG systems—from data indexing and retrieval interaction to content generation. The existence of these security risks essentially exposes the endogenous vulnerabilities of RAG systems under the ``Retrieval-Augmentation" architecture: excessive trust in external data sources, lack of effective verification mechanisms between components, and the reversibility of vector representations. As RAG technology penetrates critical fields such as finance, healthcare, and enterprise-level applications, a single vulnerability can often trigger systemic crises of trust collapse and data leakage. Therefore, relying solely on the robustness of the large model itself is insufficient to cope with the increasingly complex attack environment. It is urgent to construct a defense system covering data encryption, query filtering, and privacy protection. The following chapters will delve into security protection technologies against these threats, analyzing the principles and effectiveness of various defense strategies.

\section{RAG Security Protection Technologies}\label{sec4}
The previous chapter detailed the security risks faced by RAG systems, exposing vulnerabilities across the entire pipeline. Consequently, establishing a defense-in-depth system has become paramount for ensuring the practical deployment of RAG applications. This chapter categorizes protection technologies into two primary lines of defense: the input side and the output side. Specifically, it discusses data privacy enhancement and security admission mechanisms for the input side, as well as inference defense and information leakage protection technologies for the output side. Additionally, this chapter will present specific defense techniques designed to counter the various attacks targeting RAG systems.
\begin{figure}[!htbp]
    \centering
    \includegraphics[width=1\textwidth]{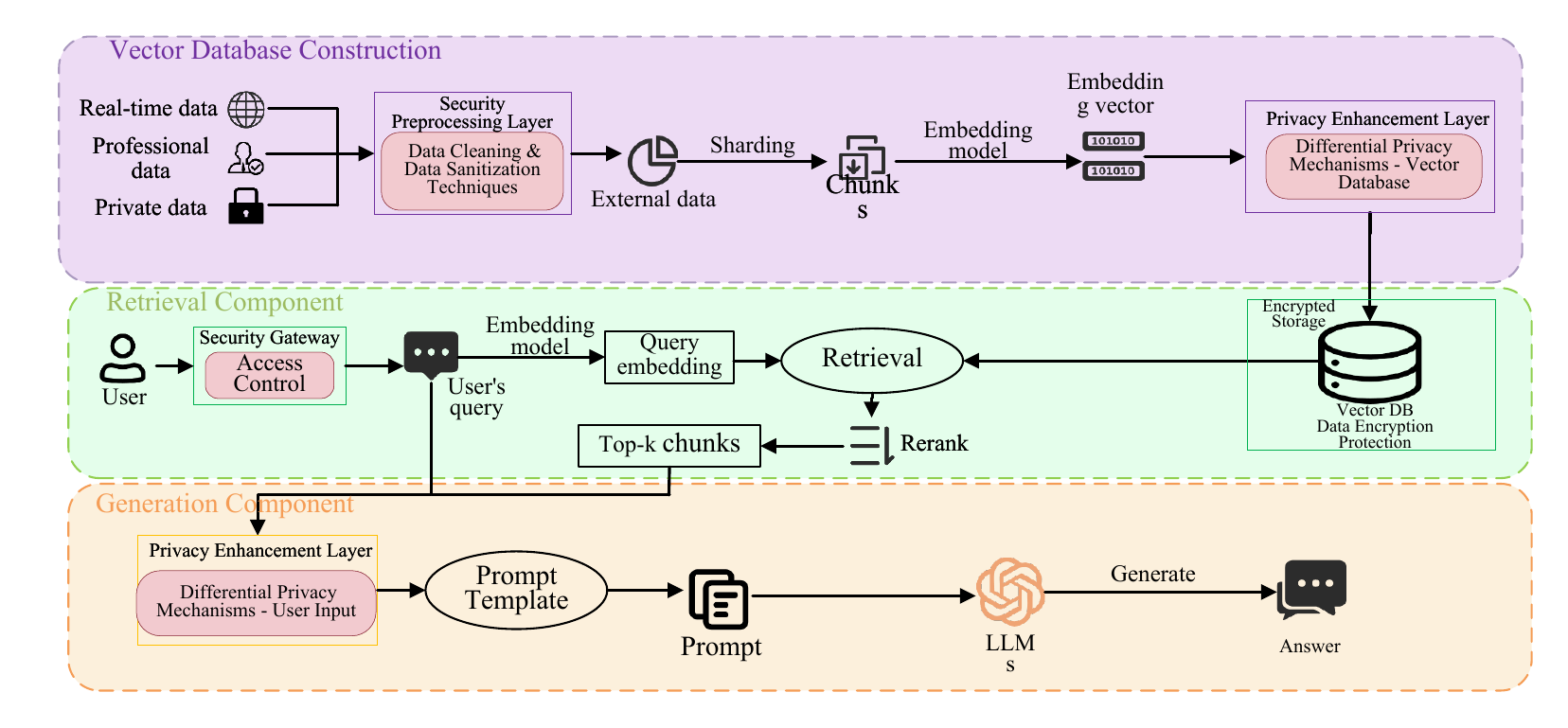}
    \caption{Pipeline-integrated defense mechanisms for secure RAG operations. The schematic maps targeted security interventions to their corresponding vulnerable nodes within the RAG architecture. Key defensive layers (pink boxes) include data sanitization and differential privacy during external data construction, database encryption and strict access control during retrieval, and input-level privacy enhancements prior to LLM generation. This end-to-end framework provides a structural blueprint for securing RAG systems against both external adversarial threats and internal data leakage.}
    \label{fig:rag_protection_tech_ecosystem}
\end{figure}
\subsection{Data Privacy Enhancement and Security Admission }\label{subsec:rag_data_privacy_enhancement}
\begin{figure}[!htbp]
    \centering
    \includegraphics[width=1\textwidth]{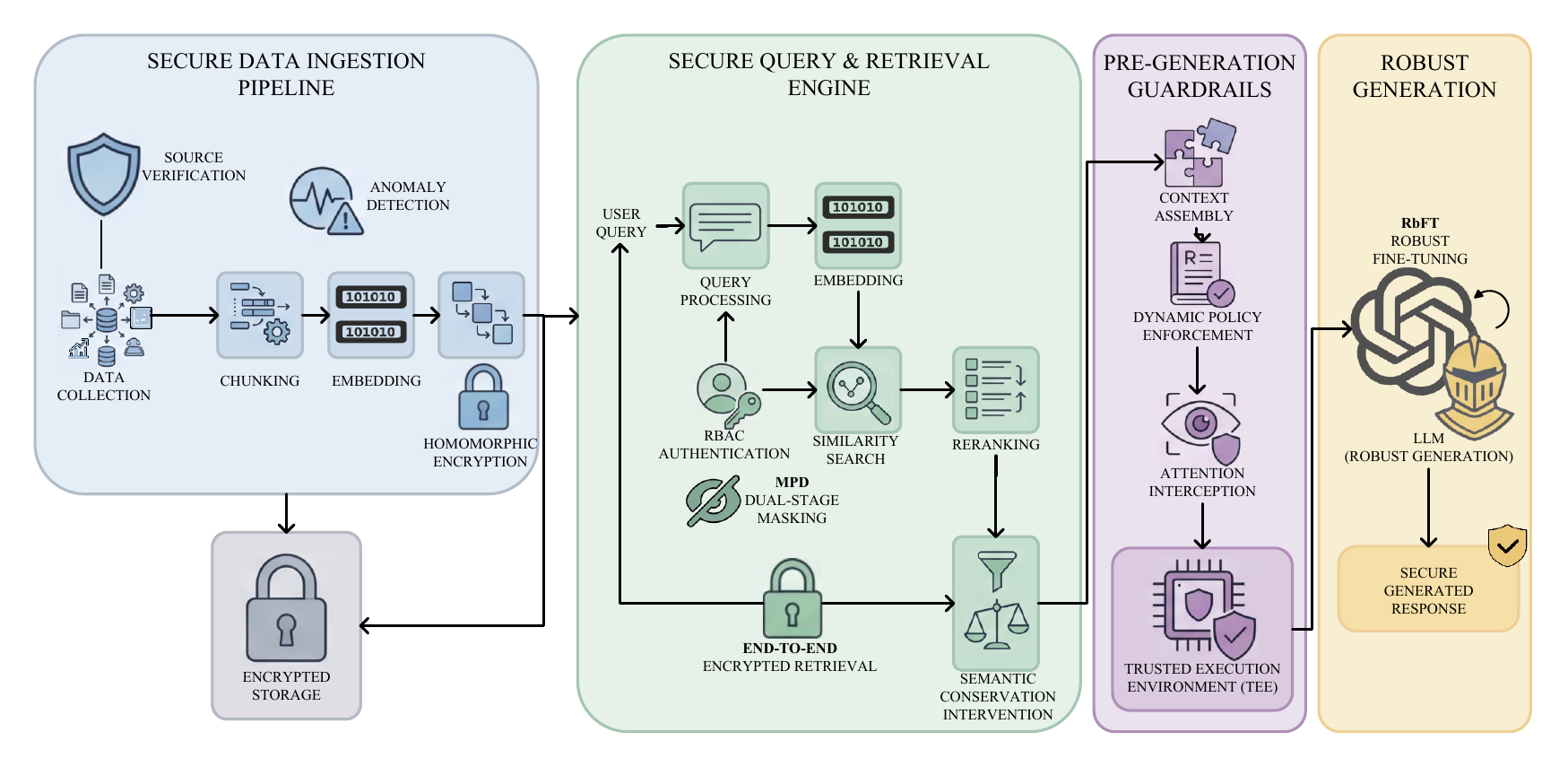}
    \caption{The proposed input-side defense pipeline for securing RAG architectures. The schematic focuses exclusively on mechanisms that protect the integrity of data entering the LLM. It maps out a four-stage input security flow: (1) securing the ingestion of external databases, (2) establishing an encrypted and authenticated query retrieval engine, and (3) deploying strict pre-generation guardrails (TEE) to intercept malicious attention manipulation. Ultimately, these stringent input-side controls provide a safe and robust context for the final generation phase.}
    \label{fig:rag_input}
\end{figure}

Serving as the initial line of defense, data privacy enhancement and security admission mechanisms at the input side constitute the foundation of overall RAG security. During the data ingestion and retrieval stages, RAG systems encounter severe security challenges. These primarily manifest as data poisoning attacks that aim to corrupt knowledge sources \citep{45,46,47,48} and adversarial attacks designed to manipulate retrieval results \citep{75,78}. If malicious texts breach the input boundary and enter the vector database or the retrieval pipeline, they directly compromise the purity of internal knowledge, subsequently degrading the stability and reliability of the generated outputs \citep{45}. Consequently, defenses at this level focus on establishing strict admission controls and preprocessing barriers. As illustrated in Figure \ref{fig:rag_input}, these measures block malicious text injection at the source and ensure consistent model predictions despite input perturbations. Current defense strategies for the input side primarily encompass several categories.
\subsubsection{Evolution from RBAC to Fine-Grained Dynamic Filtering in Access Control}\label{subsubsec:rag_access_control_evolution}
Access control constitutes the most fundamental and critical first line of defense in the RAG security architecture. Its core function is to define system boundaries, preventing unauthorized access and malicious text injection from external attackers. Concurrently, it ensures that internal sensitive data remains protected from unauthorized access or leakage, effectively mitigating various threats including data poisoning and membership inference attacks. Practical applications of traditional RAG architectures typically employ role-based access control (RBAC). For instance, \citep{113} propose a secure isolation deployment scheme for RAG systems that strictly adheres to Salesforce standards. This approach integrates RBAC with field-level security protocols to explicitly define read and write permissions for different user roles regarding specific documents or data fields within the knowledge base. Through identity authentication, this static defense mechanism intercepts unauthorized users before they initiate retrieval requests, fundamentally reducing the risk of knowledge base corruption or data theft at the source.

However, as the reasoning capabilities of Large Language Models (LLMs) enhance, simple static RBAC mechanisms are gradually revealing limitations when facing complex RAG interaction scenarios. The increasingly sophisticated semantic understanding and associative reasoning abilities of LLMs enable them to infer sensitive information from seemingly harmless non-sensitive fragments or bypass document-level permission restrictions through cross-document information aggregation. Furthermore, traditional access control struggles to defend against Indirect Prompt Injection attacks—even if the attacker does not possess high privileges, if the retrieved ``low-classification" documents contain malicious instructions planted by the attacker, the model may still be induced to execute unauthorized operations. This implies that verifying user identity solely before retrieval is no longer sufficient to counter threats, as the model's generation behavior is difficult to constrain by traditional permission rules once it acquires the context.

Addressing these challenges, novel defense strategies are evolving toward context-aware fine-grained access control. This strategy no longer relies solely on the user's static identity but incorporates dynamic permission adjudication based on the user's current query intent, contextual situation, and the sensitivity classification of retrieved content. To this end, \cite{85} proposed an innovative security protocol that embeds a Policy Enforcement Point (PEP) within the RAG information flow chain. This mechanism implements entity-level filtering and permission re-verification before retrieved content is input into the LLM. It can identify and strip sensitive entities (such as Personally Identifiable Information (PII) or trade secrets) from documents that exceed the user's permissions, ensuring the model only ``sees" information fragments the user is authorized to know. In summary, access control in RAG systems must consider the characteristics of large models, shifting from coarse-grained document-level control to fine-grained dynamic supervision throughout the entire query and generation lifecycle.

\subsubsection{Evolution from Static Storage Encryption to Homomorphic Encryption Computing in Data Protection}\label{subsubsec:rag_encryption_evolution}
Input-side protection based on encryption is a significant branch of RAG security research. Its core value lies in introducing ``unreadability" into the RAG data link , to defend against embedding inversion attacks: knowledge base texts, embedding vectors, and query contents are stored and processed in encrypted form as much as possible, thereby reducing risks associated with leaks from cloud/third-party components, operations personnel, or logs. Among these techniques, Homomorphic Encryption (HE) provides a unique pathway for ``similarity retrieval in the encrypted domain" due to its ability to support direct computation on ciphertext. It holds the promise of completing recall without exposing the plaintext of queries or vector databases, thus offering stronger end-to-end privacy guarantees in highly sensitive scenarios such as healthcare and finance.

However, encryption schemes present obvious engineering and systemic bottlenecks when implemented in RAG. First, homomorphic encryption often incurs significant computational and communication overhead, easily becoming a performance bottleneck under large-scale vector databases and high-concurrency retrieval demands, affecting retrieval latency and throughput. Second, key management, access authorization, encrypted index maintenance, and dynamic updates significantly increase system complexity. Third, while encryption primarily solves the problem of ``data invisibility," it does not inherently address semantic-level risks regarding ``whether the content is malicious/will induce model unauthorized access"—even if the retrieval stage is completed in ciphertext, indirect prompt injection and other generation-side attacks remain possible once the content is decrypted and enters the context. Therefore, encryption protection requires a more refined trade-off between security strength, system complexity, and RAG functionality/performance.

To address these challenges, existing work largely advances along two routes: ``encrypted retrieval" and ``hybrid trusted boundaries." SecureRAG splits retrieval into secure search and secure document acquisition, utilizing Fully Homomorphic Encryption (FHE) to perform similarity calculations on ciphertext vectors, thereby reducing query and embedding leakage risks, albeit with high dependence on HE operators and system optimization \citep{112}. Privacy-Aware RAG emphasizes the combination of full-lifecycle encryption with access/integrity mechanisms, leaning more toward engineering usability and closed-loop end-to-end processes, but it must also face issues of cost and trusted boundary delineation brought by encrypted computing \citep{116}. Bae et al. proposed combining untrusted cloud computing with local trusted decryption/generation. By performing encrypted retrieval via a homomorphic encryption vector database on the cloud and final generation locally, this approach reduces the privacy attack surface, though it entails more complex system links and higher requirements for interaction and edge-side resources \citep{117}. Additionally, some work attempts ``lightweight alternatives" to mitigate the high overhead of homomorphic encryption. For example, PRESS reduces privacy leakage risks during the retrieval stage through embedding space transformation with minimal performance cost, though its security is more empirical and typically difficult to achieve cryptographic-strength guarantees \citep{118}. Overall, the research focus of encryption routes is shifting from ``whether encrypted retrieval is possible" to ``achieving deployable, scalable, and composable end-to-end private RAG at acceptable costs."

In summary, protection on the RAG system input side primarily involves three technical routes: access control, data cleaning and filtering, and data encryption. The development of input-side security protection technologies indicates that the advancement of RAG defense technologies requires specific modifications to defense techniques tailored to the specific characteristics of RAG vector retrieval and large model generation, balancing defense effectiveness with the functional normality of the RAG system.
\subsubsection{Data Cleansing and Poisoning Filtration Mechanisms for Input-Side Integrity}\label{subsubsec:rag_Data_Cleansing}
Implementing rigorous cleansing and filtration during the data ingestion and query processing stages constitutes the first line of defense to ensure knowledge base purity and block data poisoning attacks. Traditional input defenses primarily rely on statistical analysis of ingested texts and semantic perturbation of user queries. In the early stages of RAG data poisoning, malicious texts typically exhibit syntactic confusion or semantic incoherence. Based on this characteristic, \citep{115} propose using perplexity as a pre-ingestion detection metric, screening potential attack payloads by identifying and discarding abnormal text segments with high perplexity. Regarding user queries, \citep{104} propose an input defense strategy based on paraphrasing. This method rewrites the initial user query to disrupt the similarity mapping between the predefined triggers of the attacker and the malicious texts in the database, thereby reducing the recall probability of malicious documents during the pre-retrieval stage.

As the construction techniques for malicious texts evolve rapidly, the threats confronting RAG systems at the input side become increasingly stealthy. \citep{46} point out that attackers leverage the generation capabilities of large language models to construct malicious content that is highly fluent, logically coherent, and statistically indistinguishable from normal text. Such stealthy poisoned data renders perplexity-based pre-cleansing mechanisms largely ineffective. Concurrently, because modern RAG systems widely employ high-precision semantic retrieval, query paraphrasing techniques struggle to sever the recall chain of malicious texts. Conventional input filtering methods typically operate at the superficial text level and fail to detect carefully disguised attack payloads, consequently eroding the defense boundaries of RAG systems during the data inflow stage.

To address the failure of traditional input filtering mechanisms, next-generation defense strategies evolve towards deep semantic validation and dynamic filtering applied after retrieval and prior to generation. During the data ingestion and index construction stages,  \citep{82} propose cross-referencing external data with authoritative sources and isolating outliers that deviate from normal semantic clusters within the embedding space, thereby blocking the input of contaminated data at the source. During the input assembly stage where retrieval results return to the language model, \citep{50} analyze the functional mechanism of malicious texts. They note that successful manipulation requires the malicious text to capture extremely high attention weights within the model. Accordingly, they propose an abnormal attention detection mechanism that monitors the attention distribution of the input context during early inference stages, dynamically identifying and filtering out retrieved segments that exhibit abnormally strong manipulative properties. In summary, effective input defenses for RAG systems can no longer rely solely on simple rule matching; instead, they require a full-chain deep filtering network encompassing data ingestion cleansing, query processing, and context validation.

\subsubsection{Detection and Pre-Defense Strategies Against Adversarial Inputs}\label{subsubsec:rag_Adversarial_Inputs}
Defense technologies against adversarial attacks in RAG systems are evolving from singular input text purification to full-chain robustness enhancement. As adversarial samples progress from simple character perturbations to complex corpus poisoning and retrieval manipulation, traditional defenses struggle to address deep logical traps and counterfactual deceptions. Consequently, the research community is exploring multi-dimensional defense paradigms. These paradigms encompass model-agnostic input certification, fine-tuning for intrinsic noise resistance in large language models, graph-based semantic reranking, and system-level threat governance. The objective is to construct a robust RAG ecosystem capable of resisting external malicious injections while performing self-correction.

Early explorations of input text purification include the masking and purifying defense framework proposed by  \citep{132}, which provides a theoretical guarantee for the security of queries and retrieved texts. Targeting character-level and word-level perturbation attacks, this research constructs a model-agnostic certified defense scheme characterized by a two-stage input processing mechanism of masking for denoising and purifying for restoration. The masker module randomly masks the input text based on rules to filter perturbations. The purifier module utilizes an improved BERT-MLM architecture to accurately restore the clean input text through soft embeddings and self-supervised fine-tuning. This method achieves effective interception under high masking rates without modifying the structure of the target generation model, providing fundamental assurance for the reliability of user queries and retrieved documents in adversarial environments.

To advance the defense frontier, recent studies delve into the foundational level of the RAG input engine, specifically the retriever mechanism. \citep{140} propose a dual defense framework comprising RAGPart and RAGMask, focusing on directly blocking the recall of adversarial corpora during the retrieval stage. This scheme leverages the segment semantic conservation property of dense retrievers for preemptive intervention. RAGPart mitigates the impact of contamination through independent embedding and combinational averaging of document segments, whereas RAGMask identifies and suppresses documents driven by malicious tokens via targeted masking. This strategy of directly intervening in the retrieval process at the input side overcomes the reliance on passive defenses during the generation stage, verifying the efficiency of achieving robust defense through retrieval purification in resource-constrained scenarios.

During the input assembly and reranking stage following retrieval recall, the GRADA framework proposed by  \citep{135} introduces a graph-based reranking mechanism for precise filtering based on semantic consistency. This method exploits the differences in semantic coherence between adversarial and benign documents to construct a document similarity weighted graph, propagating scores via a PageRank-like algorithm. This mechanism effectively clusters benign documents and suppresses isolated adversarial documents, thereby eliminating potential toxins during the reranking stage before feeding the data into the language model. GRADA introduces graph structural analysis to RAG pre-defenses for the first time, significantly reducing the success rate of input attacks that utilize semantic camouflage without sacrificing retrieval quality.

When highly stealthy adversarial inputs bypass pre-retrieval and reranking, endowing the language model with immunity to malicious inputs constitutes the final line of defense on the input side. The robust fine-tuning method proposed by \citep{133} enhances the discrimination capability of the language model when processing noisy and counterfactual retrieved inputs. Through dual-task fine-tuning for defect detection and effective information extraction, this scheme employs LoRA technology to train the model to isolate noise even when the assembled input prompt contains malicious deceptions. This approach of reducing the absolute trust of the model in malicious retrieved inputs establishes a new paradigm for improving input fault tolerance via fine-tuning.

To address increasingly complex input-level threats, \citep{134} construct a structured risk mitigation framework that elevates the perspective of input defense to full-lifecycle security governance. Based on the AI security pyramid of pain and MITRE CWE standards, this research establishes a systematic scheme encompassing threat modeling and control deployment. By accurately identifying high-risk input vectors such as prompt injection and data contamination, and by specifically deploying multi-level access and filtering controls, this framework successfully downgrades the front-end security risks confronting RAG systems to a controllable state. This provides a governance blueprint for enterprise-level RAG input defenses that balances theoretical depth with practical operability.

Overall, input-side protection for RAG systems primarily involves three technical routes, specifically access control, data cleansing and filtering, and data encryption. The evolution of these input-side security technologies indicates that advancing RAG defenses requires specific adaptations tailored to the characteristics of vector retrieval and language model generation, thereby balancing defense effectiveness with the functional integrity of the system.

\subsection{Inference Defense and Information Leakage Protection }\label{subsec:rag_inference_defense}
\begin{figure}[!htbp]
    \centering
    \includegraphics[width=1\textwidth]{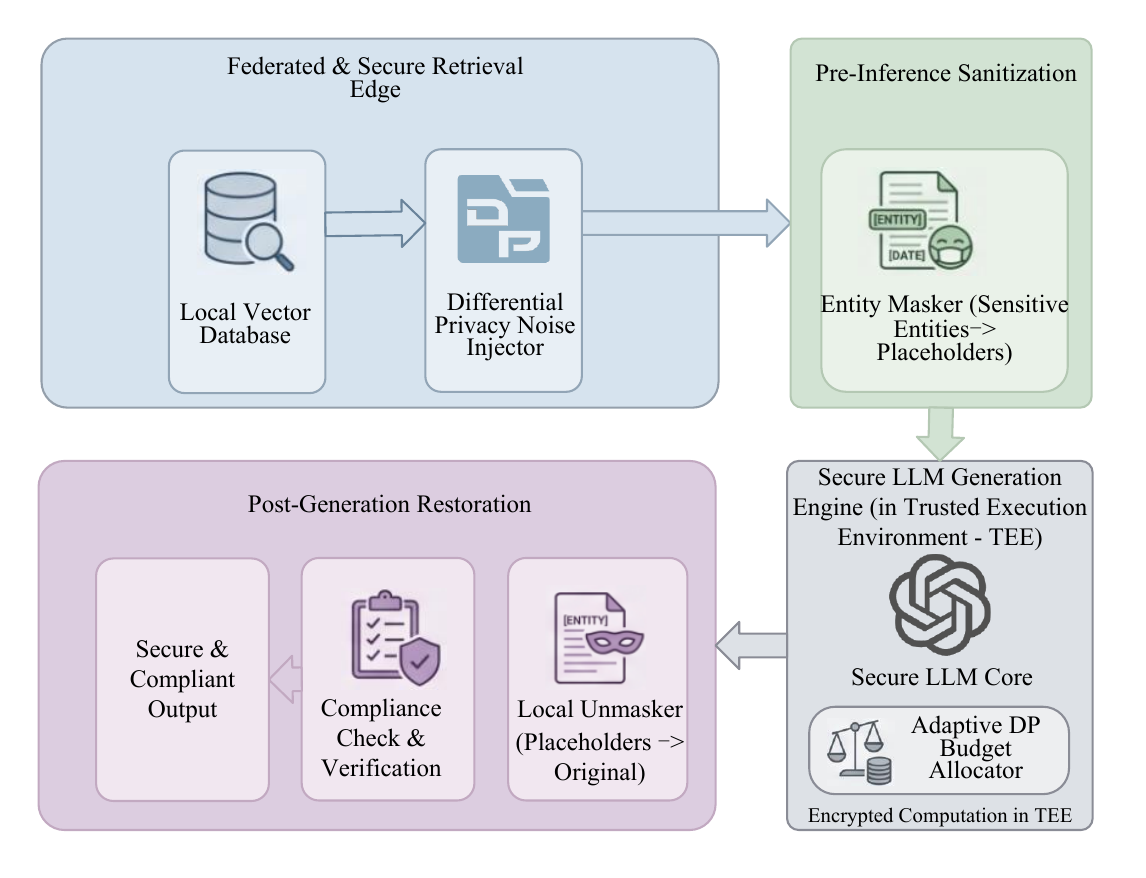}
    \caption{The proposed secure RAG system for preventing output-side privacy leakage. This diagram depicts an end-to-end privacy preservation strategy that physically and cryptographically isolates sensitive user data from the LLM. By executing retrieval and DP noise injection locally (blue), and substituting sensitive entities with placeholders (green), the system ensures the LLM generates responses based entirely on sanitized inputs within a secure enclave (TEE). The defining feature of this output-centric defense lies in the final stage (purple): the generated response, containing only placeholders, is sent back to the local client for structural unmasking and compliance verification. This design strictly confines raw sensitive information to the trusted edge, effectively neutralizing risks such as membership inference and unintended data memorization in the final output.}
    \label{fig:rag_output}
\end{figure}

Building upon solid input-side defenses, inference defense and information leakage protection technologies targeting the output side are equally crucial for safeguarding the full lifecycle of RAG. With the widespread deployment of RAG systems and the continuous expansion of knowledge bases, systems frequently process massive amounts of highly sensitive user data and private domain knowledge. In this context, any data leakage at the output end caused by model overfitting, inference attacks, or unauthorized access will severely damage user privacy and system credibility. To address this challenge, academia and industry are committed to constructing multi-layer defense frameworks integrating cryptography and distributed computing, aiming to effectively balance data utility and privacy security , as Figure \ref{fig:rag_output} . This section focuses on key privacy protection technologies adopted by RAG systems for output-side defense.
\subsubsection{Differential Privacy from Global Noise Injection to Entity-Level Fine-Grained Perturbations}\label{subsubsec:rag_differential_privacy}
Differential privacy serves as a privacy protection model with a rigorous mathematical definition. Its core concept involves adding controllable noise to data or query results to mask the contribution of specific individual data, ensuring that the addition or removal of a single sample does not significantly alter the overall output distribution. When constructing privacy-preserving RAG systems, differential privacy can be utilized to defend against membership inference and adversarial attacks. It is widely applied across two critical stages, specifically retrieval and generation. During the retrieval stage, adding noise to query embeddings prevents attackers from reverse-engineering the original user intent. In the generation stage, privacy constraints are imposed on the output text to mitigate the leakage of sensitive information from original training data or private documents. Although differential privacy provides strong theoretical security guarantees, the introduced noise inevitably degrades model utility. Balancing system accuracy and privacy under a limited privacy budget remains a core challenge in current research, particularly when processing long text generation and fine-grained analysis.

Targeting full-pipeline privacy protection for RAG systems, Grislain et al. proposed the DP-RAG framework \citep{119} to address privacy leakage from sensitive documents. This framework comprises two key components: first, document retrieval based on the Exponential Mechanism, which associates documents with unique privacy units and sets a utility function threshold \(\tau\) to ensure the retrieval process satisfies DP; second, a generation mechanism based on DP in-context learning, which generates independent augmented queries for each retrieved document and aggregates token distributions, maintaining the relevance of generated content while ensuring privacy. Experiments show that this method performs excellently in scenarios with high document redundancy (e.g., healthcare, where the same information exists in at least 100 documents). Koga et al. focused more on balancing user query privacy and long-text generation, proposing the DPSparseVoteRAG algorithm \citep{120}. Targeting sensitive external corpora, this study utilizes sparse vector technology to optimize privacy budget allocation, solving the challenge of generating long and accurate answers under limited budgets and ensuring system deployment complies with privacy regulations and data ethics.

To address the risk of verbatim leakage in the post-retrieval generation stage, several studies focus on optimizing decoding strategies. The PAD (Privacy-Aware Decoding) framework \citep{121} addresses the issue where greedy decoding might directly output sensitive content by proposing a lightweight inference-time defense mechanism. It requires no modification to the retriever or retraining of the model; by adaptively injecting calibrated Gaussian noise into token logits, it achieves privacy protection without significantly sacrificing generation quality, making it suitable for rapid deployment. Building on this, the INVISIBLEINK framework \citep{122} further optimized for long-text generation scenarios. It introduces the DClip mechanism, capable of isolating and clipping only the logit differences caused by sensitive documents, ensuring that the presence of a single sensitive document does not significantly affect the generation distribution. Compared to PAD, INVISIBLEINK achieves finer control at the logit processing and vocabulary selection levels, effectively reducing privacy budget consumption and balancing high utility with low computational cost.

Beyond general noise injection, some research explores finer-grained defense methods. He et al. proposed the LPRAG (Locally Private RAG) framework \citep{123}, based on the concept of Local Differential Privacy (LDP). Instead of crudely processing entire text segments, it identifies private entities such as words, numbers, and phrases, applying exclusive perturbations to different entity types based on an adaptive budget allocation strategy. This precise entity-level protection avoids global information distortion, ensuring robust protection for highly sensitive information like nursing data. Furthermore, Yao and Li approached defense from the vector space, proposing a method based on Random Projection \citep{124}. This method uses a Gaussian matrix to project query and document embeddings into a lower-dimensional space. While preserving the similarity required for retrieval, it perturbs the original embedding values, making it difficult for attackers to extract original information via embedding inversion. This method is applicable to KNN-LM \citep{125} and direct prompt-based RAG architectures, achieving record-level differential privacy protection.

Comparing these technologies horizontally, DP-RAG is suitable for centralized scenarios with high document redundancy, offering full-process guarantees; DPSparseVoteRAG and INVISIBLEINK excel in balancing utility for long-text generation; PAD is suitable for rapid hardening of existing systems due to its lightweight nature; while LPRAG and Random Projection provide finer solutions at the entity level and embedding layer, respectively.

\subsubsection{Output-Protected RAG under Federated Learning Frameworks}\label{subsubsec:rag_federated_learning_defense}
Federated Learning (FL), as a distributed machine learning paradigm, offers the core advantage of collaborative training where data remains stationary while the model moves. In RAG output-side defense, FL provides a viable solution: by decoupling retrieval and generation processes, it ensures the centralized generation model cannot directly access local raw data, physically blocking the risk of the generator memorizing and accidentally outputting sensitive data. This mechanism transforms traditional centralized data leakage risks into controllable model parameter exchange risks, emerges as a key technology for defending against membership inference attacks.

Addressing architectural security in the inference output stage of Federated RAG, \cite{126} constructed a comprehensive defense system aimed at eliminating privacy hazards in model outputs within highly sensitive fields like healthcare and finance. This study not only utilizes encryption protocols to protect model updates but, more critically, introduces Trusted Execution Environments (TEE) and Selective Query Routing during the inference phase. This design ensures that when the system responds to user requests, it invokes only necessary localized augmented knowledge, preventing the inference results from reverse-exposing the structure of source-side sensitive databases by minimizing data egress paths.

To further reduce information leakage risks during inference, Addison et al. proposed the C-FedRAG system \citep{127}. This system introduces Confidential Computing into the federated inference flow, designing an orchestrator running in an isolated environment. Under this architecture, document embedding retrieval is completed locally, while the aggregated inference responsible for final content generation is encapsulated within a confidential computing black box. This ``retrieval-generation" separation mechanism ensures that the final output of the RAG system is generated from securely aggregated global knowledge, preventing attackers from reverse-reconstructing local private data of participants by analyzing system outputs.

At the algorithmic level, to prevent data reconstruction caused by model parameter leakage, the FedE4RAG framework proposed by \cite{91} combines Federated Knowledge Distillation (KD-GLE module) with Homomorphic Encryption (FED-HE module). Unlike direct gradient sharing, this framework utilizes ``teacher models" produced by local trusted retrievers to guide the learning of a global ``student model." This approach ensures the final user-facing generation model learns only the generalized global knowledge distribution without memorizing specific local sensitive samples, effectively defending against membership inference attacks targeting the output model while maintaining generation quality.

As RAG technology evolves toward multi-agent collaboration, dynamic leakage during inference interaction becomes a new defense focus. The Federated Multi-Agent System (Federated MAS) proposed by \cite{128} focuses on real-time output control during the inference stage. This research designed embedded privacy-enhancing agents acting as inspection mechanisms for RAG retrieval and context interaction. When multiple agents collaborate on complex tasks, these agents can filter non-task-related redundant information in real-time, allowing only desensitized necessary information to be passed as output to other agents, thereby minimizing the data exposure surface during collaborative inference.

In domain-specific application validation, Karamanlıoğlu et al. developed a clinical decision support system integrated with stream analytics \citep{129}. This system utilizes federated RAG mechanisms to achieve inference isolation, combined with differential privacy technology, ensuring that diagnostic suggestions output by the system reflect only collective medical patterns while strictly stripping away individual patient characteristics. Such strict constraints at the output end enable the system to meet rigorous data egress requirements of regulations like GDPR and HIPAA.

In summary, RAG protection technologies under federated learning frameworks essentially reconstruct the system's output boundaries through distributed architecture. Whether it is hardware-level inference isolation in C-FedRAG, algorithm-level knowledge distillation in FedE4RAG, or interaction-level dynamic filtering in Federated MAS, the core objective is to ensure that every output generated by the RAG system undergoes strict de-identification and aggregation processing. These technologies effectively build a solid defense line preventing raw data leakage through model outputs while ensuring knowledge augmentation effectiveness.

\subsubsection{Data Sanitization via Lightweight Entity Masking and Inference-Time Privacy Isolation}\label{subsubsec:rag_data_sanitization}

Among the output-side protection technologies for RAG systems, data masking provides a lightweight and efficient reasoning isolation scheme to defend against membership inference attacks. Unlike traditional input cleansing, the core value of this technology from the perspective of output protection lies in achieving the physical decoupling of logical generation and information filling. By mapping sensitive entities, such as names and medical records, to meaningless placeholders like $<$PERSON\_1$>$, this technology compels the generator to reason exclusively at the semantic logic level without accessing or outputting actual sensitive values. This mechanism ensures that the raw response generated by the model remains fundamentally desensitized. Even if the model is manipulated by attackers or experiences hallucinations, it solely outputs secure placeholders, thereby isolating semantics from question-answering at the output end \citep{109}.

In practical protection implementation, the ``Sanitization-Inference-Restoration" workflow proposed by \cite{113} demonstrates how to defend against information leakage by controlling the output construction process:
\begin{enumerate}
    \item Output Constraints during Inference: Before inputting prompts into the large model, they undergo masking processing where sensitive terms are replaced with meaningless masks, and sensitive information along with mask positions are saved in a bidirectional mapping table. The generative model is restricted to generating responses within a safe vocabulary space containing only placeholders. This implies that the intermediate results output by the model inherently contain no user privacy, thoroughly avoiding the risk of plaintext output theft by external model service providers or man-in-the-middle attacks.
    \item Localized Output Reconstruction: The actual output process is shifted to a controlled local secure environment for execution. Based on the preset bidirectional mapping table, the system reverse-restores the response templates containing placeholders generated by the model.
\end{enumerate}

This strategy limits the role of the Large Language Model (LLM) to an untrusted logical processing engine rather than the final publisher of information. By consolidating the synthesis authority of the final output locally, this technology effectively defends against eavesdropping and reverse deduction targeting the model output end, ensuring that complete sensitive information is presented to authorized users only after local security verification.

Overall, RAG security research is transitioning from early single-threat identification and vulnerability patching to constructing systematic comprehensive defense architectures with formal guarantees. the current ecosystem of protection technologies has become increasingly comprehensive: from basic access control and data cleaning to mathematically grounded differential privacy and homomorphic encryption, and further to federated learning solving data silos and lightweight data sanitization. These technologies do not exist in isolation but are gradually forming a complementary defense-in-depth architecture—safeguarding underlying confidentiality via encryption technologies while balancing upper-layer application utility and privacy using differential privacy and sanitization technologies. This multi-layered integrated defense strategy marks the progression of RAG security research toward a more mature and trusted new stage.

Overall, research on RAG security is transitioning from early singular threat identification and vulnerability patching to the construction of systematic, comprehensive defense architectures with formal guarantees. As illustrated in Figure 8, the current ecosystem of protection technologies has become increasingly comprehensive. These technologies span from fundamental access control and data cleansing to mathematically grounded differential privacy and homomorphic encryption, further extending to federated learning for addressing data silos and lightweight data masking. These technologies are not isolated but gradually form a complementary, defense-in-depth architecture. Encryption technologies ensure underlying confidentiality, while differential privacy and masking technologies balance the usability and privacy of upper-layer applications. This multi-layered, integrated defense strategy signifies that RAG security research is advancing towards a more mature and trustworthy phase.

\subsection{Chapter Summary}\label{subsec:rag_chapter_summary_4}
In summary, defense technologies for RAG systems have constructed a defense-in-depth system covering data sources, inference pipelines, and output terminals. Addressing data poisoning threats, data cleaning and filtering mechanisms establish a trusted foundation for the knowledge base during the index construction phase through strict admission screening and anomaly detection. Facing complex adversarial attacks, robustness enhancement strategies integrate technologies such as input purification, model fine-tuning, and graph-based semantic re-ranking, significantly improving the system's logical error-correction capabilities under noise interference and counterfactual misleading. Regarding defense against membership inference attacks, privacy protection mechanisms effectively block leakage paths that utilize output distributions to reverse-engineer training data by employing differential privacy noise injection and adaptive decoding intervention. Overall, existing defense paradigms are evolving from early single-point static protection to full-lifecycle, semantic-aware, and model-agnostic dynamic collaboration, aiming to ensure that Retrieval-Augmented Generation technology achieves controllable security risks and firmly guarded privacy boundaries while maintaining high utility.
\section{Security Evaluation Standards for RAG Technology}\label{sec5}

With the development of protection technologies for Retrieval-Augmented Generation (RAG), its security evaluation has advanced from early qualitative analysis toward standardized, multi-dimensional quantitative benchmarks. Early evaluations largely adopted adversarial testing metrics from Large Language Models (LLMs). However, since RAG architectures introduce external knowledge retrieval and context fusion mechanisms, traditional single-evaluation systems struggle to cover novel threats such as embedding inversion and data poisoning. In recent years, the academic community has gradually built comprehensive evaluation frameworks covering attack success rates, retrieval fairness, generation robustness, and defense effectiveness. From baseline defense tests targeting aligned models in 2023 to specialized benchmarks for RAG-specific attacks emerging in 2025, security evaluation standards are evolving toward finer granularity, scenario-specific adaptation, and systematization.

\subsection{Multi-Dimensional Evaluation Metric System}\label{subsec:rag_evaluation_metrics}
Existing evaluation standards typically categorize security metrics into three dimensions: attack effectiveness, system utility, and detection/defense capability. Furthermore, higher-order metrics such as fairness and cognitive complexity are introduced based on research focus.

Addressing fundamental adversarial attacks, \cite{136} established Attack Success Rate (ASR) as the core metric for measuring the proportion of jailbreaks. They also introduced Perplexity and Windowed Perplexity to detect high-perplexity adversarial text. Additionally, this study emphasized the robustness-performance trade-off, asserting that enhancing defense capabilities should not excessively compromise the model's generation quality (evaluated via AlpacaEval win rates).

In complex cognitive task scenarios unique to RAG, \cite{137} introduced a fairness evaluation dimension. Its core metrics include Expected Exposure (EE-L), used to measure the deviation between the actual distribution of retrieval results and a target fair distribution (e.g., gender, geography); and Attribution Weight, which combines sentence centrality with paragraph-sentence entailment relationships to evaluate the degree of dependency of generated content on retrieval sources. Statistical tests such as the Wilcoxon Signed-Rank Test were also employed to quantify significant differences in fairness scores between the retrieval and generation ends.

Targeting component-level vulnerabilities in RAG,  \cite{138} proposed subdivided security metrics within the SafeRAG benchmark. On the retrieval side, they defined Retrieval Accuracy (RA), aiming to balance the recall of normal context with the suppression of aggressive context. On the generation side, they proposed F1 variant metrics (including F1(correct) and F1(incorrect)) to distinguish generation precision between correct and incorrect options, combining this with Attack Failure Rate (AFR) as a positive security indicator to validate defense effectiveness.

Facing large-scale poisoning threats, \cite{139} constructed comprehensive detection and robustness metrics. Beyond conventional Accuracy (ACC) and Attack Success Rate (ASR), this study specifically introduced Detection Accuracy (DACC), False Positive Rate (FPR), and False Negative Rate (FNR) targeting defense mechanisms. Simultaneously, transferability metrics were used to evaluate the generality of attacks across different RAG architectures (e.g., Multi-modal, Agent RAG), thereby revealing the shortcomings of defense systems.

\subsection{Evaluation Datasets}\label{subsec:rag_evaluation_datasets}
The construction of evaluation datasets has transitioned from general LLM security data to RAG-specific, structured constructed data.

Early research, such as \cite{136}, primarily relied on AdvBench (a set of harmful behavior prompts) and AlpacaEval (a general instruction set), focusing on assessing the model's intrinsic interference resistance. With increased attention on RAG fairness, \cite{137} selected the TREC 2022 Fair Ranking Track corpus. Based on the Anderson-Krathwohl taxonomy, they designed 8 categories of cognitive templates, deriving 368 informational queries covering understanding, analysis, and creation dimensions to test system performance under complex cognitive tasks.

Targeting RAG-specific attack vectors, \cite{138} constructed the SafeRAG dataset, the first Chinese RAG security benchmark. Based on real news text, this dataset artificially constructed ``Question-Context" pairs containing Silver Noise, inter-context conflict, soft advertisements, and white DoS attacks. It covers 5 sensitive fields including politics and finance, filling the data gap for specific bypass techniques.

To test the effectiveness of defense mechanisms in large-scale knowledge base environments, \cite{139} integrated 5 standard QA datasets, such as NQ and HotpotQA, into the RSB benchmark and constructed extended versions (EX-M, EX-L). By injecting semantically similar correct answer texts or distractors into the original datasets, they simulated the massive noise and poisoning threats faced in real retrieval environments, making the evaluation environment closer to actual deployment scenarios.

\begin{table}[htbp]
  \centering
  \caption{Comparison of RAG Evaluation Datasets}
  \label{tab:rag_eval_datasets_sorted}
  
  \tiny
  \setlength{\tabcolsep}{2pt}   
  \renewcommand{\arraystretch}{1.2} 
  
  \begin{tabularx}{\linewidth}{@{}l X X X c X@{}}
    \toprule
    \textbf{Dataset} & \textbf{Application Scopes} & \textbf{Target RAG Modules} & \textbf{Size} & \textbf{Year} & \textbf{Limitations} \\
    \midrule
    HotpotQA & Complex multi-hop logical reasoning QA tasks & Multi-hop retrieval, context fusion & 113K QA pairs (10 topics) & 2018 & Lacks security attack designs; solely evaluates foundational capabilities \\
    NQ & Single-hop knowledge QA in real-world scenarios & Retriever, generator & 307K training, 15K test pairs & 2019 & Limited to single-hop tasks; no security evaluation \\
    TREC2022Fair & 8 cognitive templates for bias verification & Retriever & 6.47M+ docs, 48 queries & 2022 & No active attack evaluations; no security threat coverage \\
    AdvBench & 32 harmful behavior-inducing attacks & Generator & 520 harmful instruction samples & 2023 & No RAG-specific attack designs; weak adaptability \\
    AlpacaEval & 10+ general instruction tasks & Generation quality & 529 general instruction samples & 2023 & No security evaluation; no RAG-specific support \\
    SafeRAG & 4 RAG-specific injection attacks & Full pipeline & 100 test pairs (5 domains) & 2025 & Small sample size; Chinese-only support \\
    \bottomrule
  \end{tabularx}
\end{table}

\begin{figure}[!htbp]
    \centering
    \adjustbox{valign=b}{\includegraphics[width=0.48\textwidth]{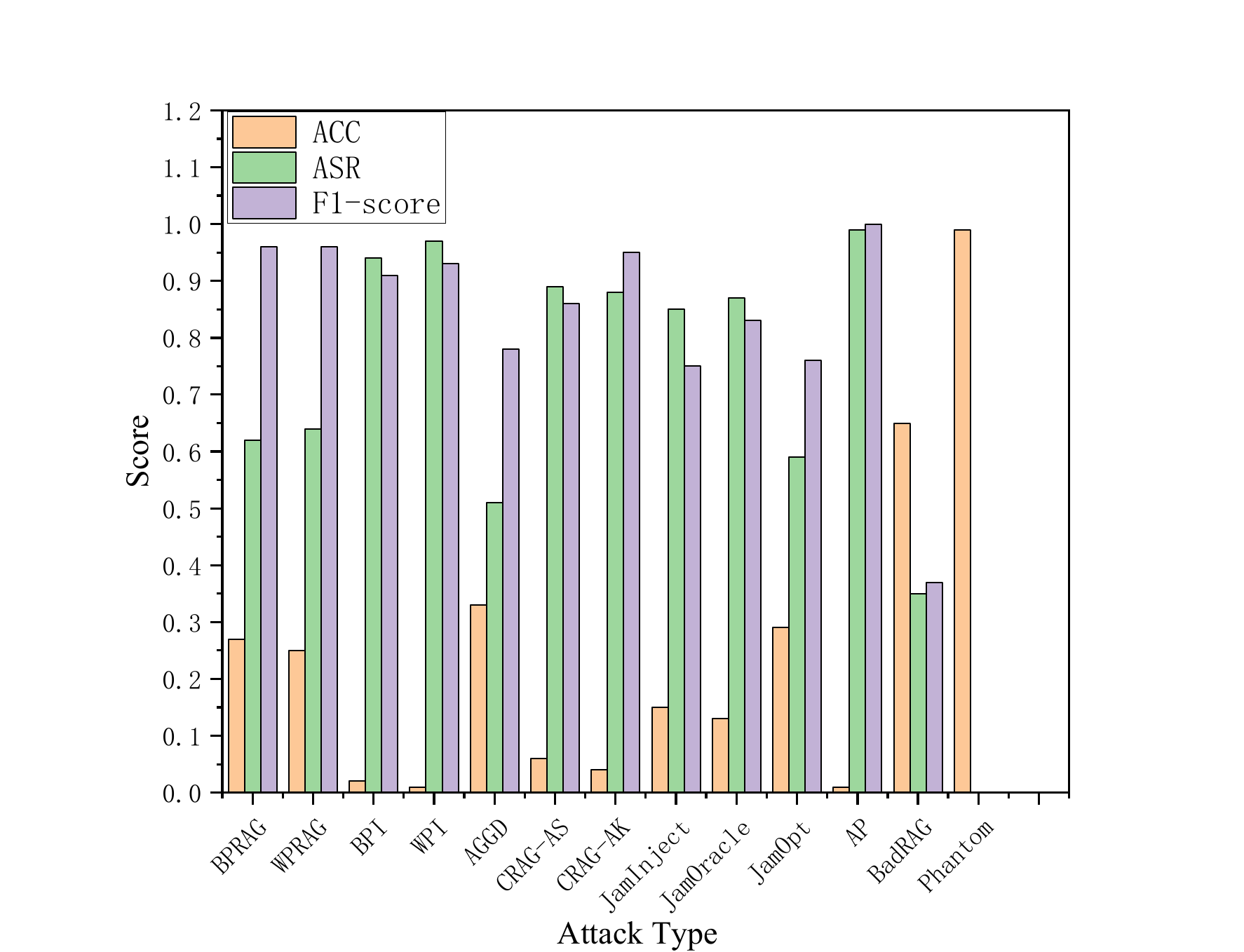}}
    \hfill
    \adjustbox{valign=b}{\includegraphics[width=0.48\textwidth]{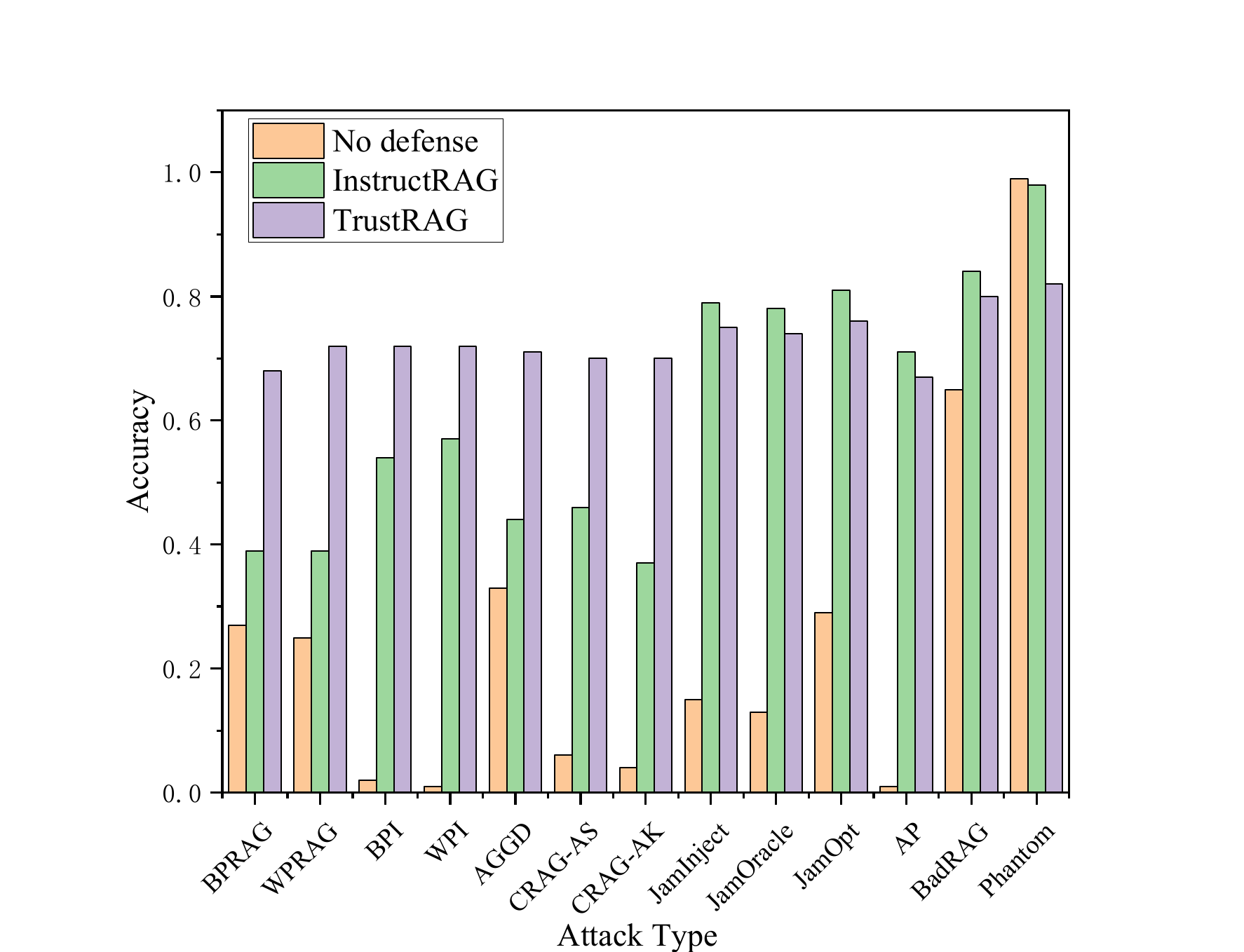}}
    
    \caption{(a) Comparative metrics of 13 attack methods evaluated on the NQ dataset. (b) Impact of InstructRAG and TrustRAG defenses on maintaining ACC against various attacks (NQ dataset)}
    \label{fig:12_13}
\end{figure}
To demonstrate the practical implications of testing in these realistic environments (e.g., the NQ dataset), a comprehensive empirical analysis of the attack-defense dynamic is required.  Figure \ref{fig:12_13} (a) quantitatively highlights the inherent fragility of undefended RAG architectures. The sharp contrast between suppressed Accuracy (ACC) and inflated Attack Success Rates (ASR) across the majority of the 13 evaluated methods underscores the lethality of modern poisoning and DoS techniques. In response to these severe vulnerabilities, Figure \ref{fig:12_13} (b) maps the recovery trajectories of the system under different protective paradigms. The data unequivocally shows that while baseline systems collapse under pressure, defense integration is crucial. Specifically, the TrustRAG framework emerges as a superior solution, significantly outperforming InstructRAG by completely neutralizing the variance between different attack vectors and maintaining a universally stable ACC baseline.

\subsection{Chapter Summary}\label{subsec:rag_chapter_summary_5}
In summary, security evaluation standards for RAG technology have formed a relatively complete system. At the metric level, there has been a multi-dimensional expansion from single Attack Success Rate to covering fairness, attribution weight, and component-level robustness. At the data level, evaluation resources have evolved from general security prompt sets to specialized datasets containing specific attack features (e.g., cognitive bias, logical conflict, large-scale noise). \cite{136} established the baseline for adversarial defense; \cite{137} and \cite{138} revealed fairness degradation and component-level vulnerabilities, respectively; while \cite{139} further validated the limitations of existing defenses under complex architectures and large-scale data. Future evaluation standards will tend more towards automation and dynamism to cope with novel security challenges brought by multi-modal fusion and Agent-based RAG systems.

\section{Summary and Future Outlook}\label{sec6}  
To clarify the developmental trajectory and overall landscape of RAG security, we systematically distill the core theoretical insights of current research, the critical challenges in practical deployment, and the primary directions for future evolution, as summarized in Table 3. This section provides an in-depth discussion based on this table. We first transcend the limitations of singular technologies to analyze the underlying logic and intrinsic flaws that contribute to the vulnerability of RAG systems. Subsequently, we address the practical barriers encountered by current defense architectures as they evolve toward dynamic and complex scenarios. Finally, from a macroscopic perspective, we propose a forward-looking technical blueprint to resolve the zero-sum trade-off between privacy and efficiency, aiming to construct a fully verifiable next-generation RAG architecture.

\begin{table}[htbp]  
    \centering  
    \caption{Insights, Challenges, and Prospects}  
    \label{tab:insights_challenges}  
      
    \scriptsize  
    \renewcommand{\arraystretch}{1.1} 
    \setlength{\tabcolsep}{3pt} 
      
    \begin{tabularx}{0.95\linewidth}{  
        >{\bfseries\raggedright\arraybackslash}p{2.2cm}  
        >{\justify\arraybackslash}m{\dimexpr0.95\linewidth-2.2cm}  
    }
    
    \toprule  
    \textbf{No.} & \textbf{Insight / Challenge / Prospect} \\  
    \midrule 
      
    Insight I & \textbf{Trust Paradox and the Contextual Security Gap shift the security focus.} The structural dependence of RAG on external retrieval causes the model to suppress internal security priors in the presence of malicious contexts (i.e., the Contextual Security Gap). This trust mechanism flaw allows minimal poisoning samples to hijack the generation process via the retrieval interface, revealing that system security is essentially bound by context integrity. \\  
      
    Insight II & \textbf{Semantic Gap renders embedding space attacks an intrinsic defect.} A fundamental inconsistency exists between human natural language logic and machine dense vector representations. Attackers exploit the statistical limitations of cosine similarity to create ``adversarial collisions.'' Without a symbolic logic verification layer, relying solely on vector retrieval cannot eradicate covert attacks targeting the embedding space. \\  
      
    Insight III & \textbf{Data flow is trapped in a Privacy-Efficiency zero-sum game.} In existing solutions, Fully Homomorphic Encryption (FHE) is inefficient, TEEs face side-channel risks, while plaintext retrieval turns RAG into a ``data leakage amplifier.'' Attackers can steal sensitive records via inference side-channels (e.g., Logits analysis) or embedding inversion, making it difficult to balance real-time performance with confidentiality. \\  
      
    \midrule  
      
    Challenge I & \textbf{Runtime encrypted data streams face dual bottlenecks of efficiency and side channels.} Static encryption cannot protect high-frequency runtime vector operations. Searchable Encryption (SE) faces a conflict between computational overhead and accuracy in high-dimensional retrieval, with complex dynamic index maintenance. Furthermore, pure encrypted retrieval fails to mask access patterns, enabling attackers to infer query intent or knowledge base distribution via traffic statistics. \\  
      
    Challenge II & \textbf{The Agentic evolution introduces risks of physical cascading failures.} As RAG transitions toward active task execution, retrieval poisoning can translate into physical malicious operations. Facing network-wide ``delusions'' caused by cross-agent infection and uncontrolled tool invocation, traditional defenses based on text matching struggle to semantically block malicious API calls disguised as normal logic. \\  
      
    Challenge III & \textbf{Evaluation benchmarks are lagging and constrained by multi-dimensional metric coupling.} Existing evaluations lack end-to-end standards, with security and efficacy metrics highly coupled. Static datasets fail to capture black-box ranking attacks or adaptive stealthy poisoning, while automated testing relying on ``LLM-as-a-Judge'' suffers from bias and hallucinations, making it difficult to quantify the defense boundaries against unknown threats. \\  
      
    \midrule  
      
    Prospect I & \textbf{Constructing an efficient confidential computing architecture to break the zero-sum game.} The future lies in developing dedicated Privacy-Enhancing Technologies (PETs), such as balancing utility and compliance via sparse Differential Privacy, and designing hybrid encryption protocols (SE+HE) for efficient ciphertext Approximate Nearest Neighbor (ANN) search, thereby resolving runtime privacy leakage while ensuring real-time performance. \\  
      
    Prospect II & \textbf{Establishing a standard for White-box RAG (SAG) based on cryptographic verification.} Systems will shift from black-box trust to provable security. By isolating poisoning via full-link encryption, utilizing recursive SNARKs for practical Zero-Knowledge data provenance, and integrating Verifiable Vector Retrieval technologies, the integrity and immutability of retrieval results can be mathematically guaranteed, building a trustless secure interaction. \\  
      
    Prospect III & \textbf{Shifting from content security to dynamic governance of agent behavior.} With the rise of Agentic RAG, defense focus must shift to ``Behavioral Security.'' Utilizing RLHF for Security to constrain tool invocation and establishing dynamic runtime sandboxes to isolate side effects are essential to prevent hijacked agents from autonomously executing destructive operations within enterprise networks. \\  
      
    \bottomrule  
    \end{tabularx}  
\end{table} 

\subsection{Insight}\label{subsec:rag_summary_1}

Following a deep investigation into specific defense technologies for RAG, we analyze the endogenous roots of its vulnerabilities from a theoretical perspective. The RAG architecture is not merely a stacking of components; by introducing external evidence to anchor the generation process, it triggers a profound \textbf{Trust Paradox}. To enhance utility, the system is forced to develop a structural dependency on externally retrieved content, and this dependency serves as the very lever exploited by attackers. Research indicates that even models subject to strict safety alignment tend to suppress their internal safety priors and credit external data when facing malicious or misleading context \citep{164}. This phenomenon is termed the \textbf{Contextual Security Gap}, shifting the center of gravity for security from model parameters to the retrieval interface. For instance, attacks such as CorruptRAG and PoisonedRAG demonstrate that injecting a minimal amount of malicious samples into a massive corpus can induce the model to generate predefined false answers or malicious instructions with a success rate often exceeding 90\% \citep{165}. This reveals the essential characteristic that RAG security is constrained by the semantic integrity of the retrieval corpus.  
  
Beyond the shift in trust mechanisms, another major endogenous defect of RAG systems stems from the \textbf{Semantic Gap}---the fundamental inconsistency between natural language logic understood by humans and dense vector representations processed by machines. Existing vector databases rely on statistical features like cosine similarity for retrieval, failing to capture true logical entailment. Attackers exploit this gap to craft adversarial samples: a string of text that appears as gibberish or benign to humans can, after encoding by an embedding model, possess a vector that perfectly overlaps with high-confidentiality data or specific trigger instructions in the embedding space. As long as RAG systems overly rely on singular dense vector representations without a symbolic logical verification layer, embedding inversion attacks remain an ineradicable theoretical bottleneck \citep{166}.  
  
Furthermore, at the data flow level, current RAG solutions are trapped in a severe \textbf{``Privacy-Efficiency'' Zero-Sum Game}. Fully Homomorphic Encryption (FHE), while theoretically secure, incurs immense computational overhead, making it unsuitable for real-time interaction demands; Trusted Execution Environment (TEE) solutions are faster but are limited by hardware roots of trust and face side-channel risks. Lacking strong privacy protection, RAG systems effectively act as efficient \textbf{``Data Leakage Amplifiers.''} Since Top-$k$ documents are directly input as context, attackers can induce outputs via prompt injection or analyze logits to infer sensitive records within the database with high precision \citep{26}. Simultaneously, once vector index access is compromised, the risk of recovering original text using embedding inversion techniques renders the security of the vector database itself a weak link in the entire chain \citep{45}.  
  
Based on these underlying insights, although RAG security research has made progress in local defenses, it still faces severe challenges in the face of \textbf{Agentic} trends and zero-trust demands. Future research must move beyond ``patch-style defense'' and evolve toward full-chain verifiability and behavioral security governance.

\subsection{Challenges}\label{subsec6-1}  
  
\subsubsection{Efficiency Bottlenecks and Privacy Risks of Runtime Encrypted Data Flows}\label{subsubsec6-1-1}  
Current RAG encryption defenses are mostly limited to the static storage phase, which prevents direct theft \citep{116} but fails to protect runtime data flows. The core of RAG relies on high-frequency operations within vector databases; however, performing similarity retrieval on embedding vectors in ciphertext faces severe challenges. First is the contradiction between computational efficiency and retrieval precision: \textbf{Searchable Encryption (SE)} schemes typically incur high computational and communication overheads, making it difficult to process high-dimensional vector retrieval without sacrificing RAG real-time performance \citep{150}. Second is the complexity of index maintenance: the dynamic updates of RAG knowledge bases require encrypted indexes to support efficient addition, deletion, and modification while ensuring both forward and backward secrecy. Finally, there are deep privacy leakage risks: embedding vectors themselves may imply plaintext semantics \citep{92}, and simple encrypted retrieval cannot mask access patterns, allowing attackers to implement Membership Inference Attacks via statistical inference.  
  
\subsubsection{Compound Risks and Cascading Failures in Agentic Systems}\label{subsubsec6-1-2}  
As RAG evolves from passive QA to task-executing \textbf{Agentic RAG}, security risks escalate qualitatively. Retrieval poisoning no longer leads merely to text errors but can translate into malicious operations in the physical world. This shift introduces \textbf{cross-agent infection risks}, where a poisoned agent may propagate erroneous information to the entire agent network via a shared memory pool, causing ``systemic delusion'' \citep{167}. Simultaneously, the uncontrollability of tool usage intensifies; text-matching filters fail to judge the semantic legitimacy of API calls (e.g., modifying code, sending emails), especially when malicious instructions disguise themselves as normal business logic \citep{168}. Current static defenses lack active mechanisms like AgenticRed \citep{169} that self-evolve to block complex attack chains in real-time, leaving systems exposed to the threat of cascading failure.  
  
\subsubsection{Fragmentation of Evaluation Benchmarks and Lag in Adversarial Evolution}\label{subsubsec6-1-3}  
Current RAG security assessment lacks end-to-end standardized benchmarks, making it difficult to fairly quantify the effectiveness of defense technologies. First is the complexity of evaluation tool design: the multi-modular nature of RAG couples security highly with metrics like retrieval quality and generation faithfulness, and existing evaluation frameworks still require refinement in numerical derivation details \citep{154}. Second is the diversity and stealthiness of attacks: as novel attacks designed to bypass detection---such as stealthy poisoning---constantly emerge \citep{52}, benchmarks based on static datasets are gradually becoming ineffective. Furthermore, the ability to detect unknown threats in black-box attack scenarios is difficult to measure effectively using existing fragmented tests.  
  
\subsection{Outlook}\label{subsec6-2}  
  
\subsubsection{Constructing Efficient Full-Chain Confidential Computing Architectures}\label{subsubsec6-2-1}  
To break the privacy-efficiency zero-sum game, future research must develop high-performance privacy-enhancing technologies tailored for RAG. On one hand, the application of sparse differential privacy algorithms, such as \textbf{DPSparseVoteRAG}, should be promoted. By sparsifying vectors before noise injection, these methods retain key semantic features while meeting strict privacy budgets, achieving a balance between privacy and utility \citep{120}. On the other hand, research should focus on exploring hybrid protocols combining efficient Searchable Encryption and Homomorphic Encryption, designing index structures specifically for approximate nearest neighbor search in ciphertext vector spaces to reduce computational complexity while ensuring retrieval precision.  
  
\subsubsection{Moving Toward Cryptographically Verifiable White-Box RAG Architectures}\label{subsubsec6-2-2}  
Future RAG systems will transcend trust-based black-box models to establish \textbf{Provably Secure RAG} standards. Research focuses should center on full-chain encryption and isolation, adopting pre-stored fully encrypted schemes to ensure retrieval content remains invisible during storage, transmission, and computation, fundamentally blocking poisoning based on database access \citep{151}. Simultaneously, practical \textbf{Zero-Knowledge Provenance} should be achieved via algorithmic optimization (e.g., Recursive SNARKs), providing concise cryptographic proofs for every generation to certify that the response originates from an authenticated and untampered dataset \citep{170}. Additionally, integrating verifiable vector retrieval technologies will enable users to mathematically verify the accuracy of Top-$k$ results, eliminating retrieval hijacking risks and constructing secure interaction mechanisms without trust assumptions.  
  
\subsubsection{Behavioral Security Governance for Agentic RAG}\label{subsubsec6-2-3}  
Future RAG will no longer be a simple ``retrieval-generation'' pipeline but will evolve into \textbf{Agentic RAG}. Agents will not only retrieve information but also call tools, execute code, and even modify databases. Security research will shift from singular ``Content Security'' to ``\textbf{Behavioral Security}.'' Preventing poisoned RAG Agents from autonomously executing malicious operations within enterprise intranets (e.g., database deletion, phishing) represents a new research high ground. Future technical directions include developing Reinforcement Learning-based safety alignment algorithms specifically to constrain Agent tool-use behaviors, as well as establishing dynamic runtime sandboxes to isolate Agent side effects \citep{171}.  
  
\section{Conclusion}\label{sec7}

Retrieval-Augmented Generation (RAG) systems, through the synergy of retrieval and generation modules, effectively expand the knowledge boundaries of Large Language Models (LLMs) and improve the accuracy and timeliness of outputs. They have demonstrated extensive application potential in critical fields such as healthcare, education, and smart contracts. However, their multi-modular architecture and dependency on external knowledge introduce unique security risks distinct from traditional large models. Consequently, building a comprehensive security protection system has become a core prerequisite for the large-scale implementation of RAG technology.

This survey has systematically reviewed the current status of RAG security research. Starting from the architectural foundation, it comprehensively analyzed prevailing security threats: data poisoning attacks tamper with system outputs by injecting malicious data; membership inference attacks threaten the privacy of knowledge bases; adversarial attacks manipulate retrieval and generation processes via stealthy perturbations; and other risks such as embedding inversion and indirect prompt injection further exacerbate system vulnerabilities. These threats impact the reliability, integrity, and confidentiality of RAG systems from multiple dimensions, including data, models, and privacy.

Addressing these security challenges, existing defense technologies have coalesced into two core directions: data and model security and privacy protection. Methods such as data cleaning, encryption protection, and access control have solidified the security perimeter for data and models, while technologies like federated learning and anonymization provide effective support for privacy protection. These defense strategies offer viable pathways for mitigating currently known security risks. Simultaneously, the field has preliminarily established experimental standards and evaluation benchmarks based on specific attack types, providing crucial tools for the quantitative analysis of attack and defense effectiveness.

Finally, regarding the challenges and threats facing RAG security research, this survey provided a future outlook. It highlighted that RAG security research still faces numerous open challenges: insufficient real-time defense capabilities during system runtime, a lack of formal security proofs for existing defense technologies, and limited coverage and adaptability of benchmark tests. Future research should focus on advancing dynamic encryption defense, provably secure RAG, and comprehensive standard benchmarking. Through technical innovation and system perfection, the goal is to construct RAG systems that possess security, efficiency, and practicality. This survey aims to serve as a reference for research dedicated to promoting the security and reliability of RAG systems in their further practical development.
\nocite{51,131,108,147,157,86,90,156,158,159,148,89}


\bibliography{sn-bibliography}

\end{document}